\documentclass[journal]{IEEEtran}
\usepackage{amsmath}
\usepackage{amsfonts}
\usepackage{amssymb}
\usepackage{amsthm}
\usepackage{graphicx}
\usepackage{array}
\usepackage[para,online,flushleft]{threeparttable}
\usepackage{xcolor}

\theoremstyle{definition}
\newtheorem{definition}{Definition}
\theoremstyle{property}
\newtheorem{property}{Property}
\theoremstyle{assumption}
\newtheorem{assumption}{Assumption}
\theoremstyle{lemma}
\newtheorem{lemma}{Lemma}
\theoremstyle{theorem}
\newtheorem{theorem}{Theorem}

\ifCLASSINFOpdf
\else
\fi
\begin{document}
\title{Best Linear Approximation of Nonlinear Continuous-Time Systems Subject to Process Noise and Operating in Feedback}

\author{Rik~Pintelon,~\IEEEmembership{Fellow,~IEEE,}
        Maarten~Schoukens,~\IEEEmembership{Member,~IEEE,}
        and~John~Lataire,~\IEEEmembership{Member,~IEEE}% <-this % stops a space
\thanks{This work was supported in part by the Research Council of the Vrije Universiteit Brussel, by the Research Foundation Flanders (FWO-Vlaanderen), and by the Flemish Government (Methusalem Fund METH1). Maarten Schoukens is supported by the European Union's Horizon 2020 research and innovation programme under the Marie Sklodowska-Curie Fellowship (grant agreement nr. 798627)}
\thanks{R. Pintelon and J. Lataire are with the Department
ELEC of the Vrije Universiteit Brussel, 1050 Brussel, Belgium (email: Rik.Pintelon@vub.ac.be and John.Lataire@vub.ac.be).}% <-this % stops a space
\thanks{M. Schoukens is with the Control Systems research group of the Eindhoven University of Technology, Eindhoven, The Netherlands (email: m.schoukens@tue.nl).}% <-this % stops a space
}


\markboth{IEEE Transactions on Instrumentation and Measurement,~Vol.~xx, No.~xx, xx}%
{Shell \MakeLowercase{\textit{et al.}}: Best Linear Approximation of Nonlinear Continuous-Time Systems Subject to Process Noise and Operating in Feedback}



\maketitle

\begin{abstract}
In many engineering applications the level of nonlinear distortions in frequency response function (FRF) measurements is quantified using specially designed periodic excitation signals called random phase multisines and periodic noise. The technique is based on the concept of the best linear approximation (BLA) and it allows one to check the validity of the linear framework with a simple experiment. Although the classical BLA theory can handle measurement noise only, in most applications the noise generated by the system -- called process noise -- is the dominant noise source. Therefore, there is a need to extend the existing BLA theory to the process noise case. In this paper we study in detail the impact of the process noise on the BLA of nonlinear continuous-time systems operating in a closed loop. It is shown that the existing nonparametric estimation methods for detecting and quantifying the level of nonlinear distortions in FRF measurements are still applicable in the presence of process noise. All results are also valid for discrete-time systems and systems operating in open loop.  
\end{abstract}

\begin{IEEEkeywords}
best linear approximation, nonlinear  systems, feedback, continuous-time, process noise, nonparametric estimation, frequency response function.
\end{IEEEkeywords}

% \fi
\IEEEpeerreviewmaketitle

\section{Introduction}
\IEEEPARstart{S}{ince} most real-life systems behave -- to some extent -- nonlinearly, it is important to quantify the impact of the nonlinearties on the linear modeling framework. A powerful tool for detecting and quantifying the presence of nonlinear (NL) distortions in frequency response function measurements is the best linear approximation (BLA) introduced in \cite{Schoukens_etal_TAC_1998} for nonlinear time-invariant systems operating in open loop, and generalized in \cite{Pintelon_etal_TIM_2013a} for the closed loop case. The major limitation of the classical BLA framework is that it can handle measurement noise only \cite{EnqvistLjung_AUT_2005,Pintelon_book_2012}, while in practice the  noise generated by the system -- called process noise -- is mostly  dominant. Hence, it is important to analyze the impact of the process noise on the BLA.

\begin{figure}[t!] %[htbp]: here, top, bottom, page (follow by ! to enforce, e.g. b!)
\centering
\includegraphics[width=0.75\columnwidth]{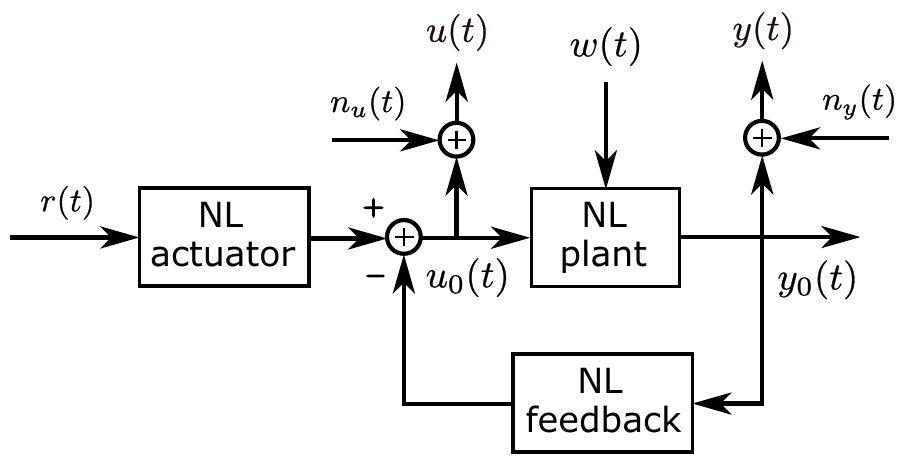} 
\caption{Noisy input $u(t)$, noisy output $y(t)$ measurement of a nonlinear (NL) time-invariant plant subject to process noise $w(t)$ and operating in closed loop. $n_u(t)$ and $n_y(t)$ are -- possibly jointly correlated -- stationary random processes that are independent of the known reference signal $r(t)$. The process noise $w(t)$ is independently distributed of the reference $r(t)$ and the input-output measurement noise $n_u(t)$ and $n_y(t)$.}
\label{fig:EIV_NLTI_PN_ClosedLoop}
\end{figure}

Beside control applications \cite{Wernholt_etal_TIM_2008,VanderMaas_etal_CEP_2016} and amplifiers operating in closed loop \cite{Pintelon_etal_TIM_2004}, feedback is present in any experimental setup where the plant is excited by a non-ideal actuator \cite{Pintelon_etal_TIM_2013b}. It emphasizes the importance of handling nonlinear systems subject to process noise and operating in a closed loop [see Figure \ref{fig:EIV_NLTI_PN_ClosedLoop}].

Using the BLA one can easily check the validity of the linear framework in practical applications such as, for example, operational amplifiers \cite{Pintelon_etal_TIM_2004}, industrial robots \cite{Wernholt_etal_TIM_2008}, bit-error-rate measurements in telecommunication \cite{Vandersteen_etal_TIM_2009}, characterization of lithium ion batteries \cite{Firouz_etal_2016}, control of a medical X-ray system \cite{VanderMaas_etal_CEP_2016}, voltage instrument transformers \cite{Faifer_etal_TIM_2018}, and current transformers \cite{Cristaldi_etal_TIM_2019}. In addition, the dependence of the BLA on the excitation power spectrum also provides some guidance for nonlinear model selection \cite{Lauwers_etal_TIM_2008,Schoukens_etal_AUT_2015}.

Recently the influence of process noise on the BLA has been studied for discrete-time Wiener-Hammerstein systems \cite{GiordanoSjoberg_CDC_2016} and for nonlinear discrete-time systems that can be approximated arbitrarily well in mean square sense by a finite degree discrete-time Volterra series \cite{Schoukens_etal_TAC_2020}. Compared with \cite{GiordanoSjoberg_CDC_2016, Schoukens_etal_TAC_2020}, the new contributions of this paper are:
\begin{itemize}
\item Nonlinear continuous-time systems are handled.
\item Additional properties of the best linear approximation and its output residual are proven.
\item The full feedback case is considered where all dynamical systems can be nonlinear and subject to process noise, and where the output as well as the input measurements are noisy [see Figure \ref{fig:EIV_NLTI_fullPN_ClosedLoop}].
\item A multiple experiment procedure is proposed to differentiate nonlinear input-output behavior from nonlinear input-process noise interactions. 
\item All results are valid for continuous-time as well as discrete-time nonlinear systems.
\item Verification of the theory on simulations (discrete-time) and real measurements (continuous-time) of nonlinear feedback systems. 
\end{itemize}

The paper is organized as follows. First, the class of excitation signals (Section \ref{sec:ClassExcitSignals}) and the class of nonlinear feedback systems (Section \ref{sec:ClassNonLinSystems}) for which the theory applies are defined. Next, the best linear approximation and its output residual are studied in detail (Section \ref{sec:BestLinearApprox}). Further, the theory is illustrated on simulations (Section \ref{sec:SimulationExample}) and real measurements (Section \ref{sec:MeasurementExample}). Finally, some conclusions are drawn (Section \ref{sec:Conclusions}).

\section{Class of Excitation Signals}
\label{sec:ClassExcitSignals}
A special class of periodic excitation signals that plays an important role in the detection and quantification of nonlinear distortions in frequency response function (FRF) measurements are random phase multisines.
\begin{definition}[Random Phase Multisine]
\label{def:RandomPhaseMultisine}
A real signal $r(t)$ is a random phase multisine if
\begin{subequations}
\label{subeq:RandomPhase_MS}
\begin{equation}
r(t)=\sum_{k=-\frac{N}{2}+1}^{\frac{N}{2}-1}R_{k}e^{j2\pi \frac{k}{N}f_{\mathrm{s}}t}
\label{eq:RandomPhase_MS}
\end{equation}
with $R_k=\overline{R}_{-k}=|R_k|e^{j\angle R_k}$, $f_{\mathrm{s}}$ the clock frequency of the arbitrary waveform generator, and $N\in \mathbb{N}$ the number of samples within one signal period. The random phases $\angle R_k \in [0,2\pi)$, $k\ne 0$, of the Fourier coefficients $R_k$ are independently (over $k$) distributed such that
\begin{equation}
\mathbb{E}\{e^{j\angle R_k}\}=0~~\mathrm{and}~~\mathbb{E}\{e^{j2\angle R_k}\}=0.
\label{eq:RandomPhaseConditions}
\end{equation}
\end{subequations}
The deterministic amplitudes of the Fourier coefficients $R_k$ are either zero (the harmonic is not excited) or satisfy $R_k=\hat{R}(kf_{\mathrm{s}}/N)/\sqrt{N}$, where the function $S_{\hat{R}\hat{R}}(f)=|\hat{R}(f)|^2$ is uniformly bounded $0\leqslant S_{\hat{R}\hat{R}}(f)\leqslant M_R <\infty$ with a finite number of discontinuities on $[0,f_{\mathrm{s}}/2]$.
\end{definition}
Note that the DC-value, $r_{\mathrm{DC}}=R_0$, of the random phase multisine \eqref{subeq:RandomPhase_MS} defines the set-point of the nonlinear system. It can have a major impact on the nonlinear distortions in the FRF measurement.

If the amplitudes of the Fourier coefficients in \eqref{eq:RandomPhase_MS} are also randomly distributed, then $r(t)$ is a periodic noise signal.

\begin{definition}[Periodic Noise]
\label{def:PeriodicNoise}
Consider the signal \eqref{eq:RandomPhase_MS} where the amplitudes $|R_k|=|\hat{R}(kf_{\mathrm{s}}/N)|/\sqrt{N}$ of the Fourier coefficients are either zero, or the realization of an independent (over $k$) random process, with $S_{\hat{R}\hat{R}}(f)=\mathbb{E}\{|\hat{R}(f)|^2\}$ a uniformly bounded function with a finite number of discontinuities on the interval $[0,f_{\mathrm{s}}/2]$. If the random phases $\angle R_k$ satisfying \eqref{eq:RandomPhaseConditions}, are independently distributed of the random amplitudes $|R_k|$, then $r(t)$ is a periodic noise signal. 
\end{definition}
The discrete Fourier transform (DFT) of random phase multisines and periodic noise signals has the following property.
\begin{property}[DFT of Random Phase Multisines and Periodic Noise Signals]
\label{prop:DFT_RPM_PNS}
The scaled discrete Fourier transform (DFT)
\begin{equation}
X(k)=\frac{\mathrm{DFT}\left\{ x(nT_{\mathrm{s}})\right\}}{\sqrt{N}} =\frac{1}{\sqrt{N}}\sum_{n=0}^{N-1}x(nT_{\mathrm{s}})e^{-j2\pi kn/N}\label{eq:DFT_sqrt_N_scaling}
\end{equation}
of $N$ samples covering one period of a random phase multisine \eqref{subeq:RandomPhase_MS} or periodic noise signal, equals
\begin{equation}
R(k)=\frac{1}{\sqrt{N}}\frac{\hat{R}(kf_{\mathrm{s}}/N)}{\sqrt{N}}N=\hat{R}(kf_{\mathrm{s}}/N)
\label{eq:RelationDFT_ScaledFourierCoeff}
\end{equation}
for $k=1,2,\ldots,N/2-1$ [proof: see Section 2.3 of \cite{Pintelon_book_2012}], where $\mathbb{E}\{|R(k)|^2\}$ is uniformly bounded [proof: $S_{\hat{R}\hat{R}}(f)=\mathbb{E}\{|\hat{R}(f)|^2\}$ is uniformly bounded; see Definitions \ref{def:RandomPhaseMultisine} and \ref{def:PeriodicNoise}].
\end{property}

According to the central limit theorem [see Theorem 27.3 of  \cite{Billingsley_book_1995}], the random phase multisine (Definition \ref{def:RandomPhaseMultisine}) and the periodic noise (Definition \ref{def:PeriodicNoise}) are -- within one signal period -- asymptotically ($N\rightarrow \infty$) normally distributed with mean value $\mathbb{E}\{r(t)\}=\mathbb{E}\{R_0\}$ and asymptotic variance $\sigma_r^2=\lim_{N\rightarrow\infty}\mathrm{var}(r(t))$
\begin{equation}
\sigma_r^2 = \frac{2}{f_{\mathrm{s}}}\int_{0}^{\frac{f_{\mathrm{s}}}{2}}S_{\hat{R}\hat{R}}(f)df
\label{eq:AsVar_MS}
\end{equation}
where $S_{\hat{R}\hat{R}}(f)=\mathbb{E}\{|\hat{R}(f)|^2\}$ [proof: see Appendix \ref{sec:ProofAsVarMultiSine}].

Although the central limit theorem indicates an asymptotic ($N\rightarrow \infty$) equivalence between, on the one hand, random multisines and periodic noise, and the other hand, Gaussian noise, their power spectral densities are fundamentally different. Indeed, stationary Gaussian noise has a continuous power spectral density, while that of a periodic signal consists of the sum of Dirac impulses. To establish an equivalence class between periodic and random signals we need the concept of Riemann equivalent power spectra \cite{Schoukens_etal_TIM_2009}.
\begin{definition}[Riemann Equivalent Power Spectra]
\label{def:RiemannEquivalentPowerSpectra}
Two stationary random and/or periodic signals $r_1(t)$ and $r_2(t)$, with respective power spectral densities $S_{r_1r_1}(j\omega)$ and $S_{r_2r_2}(j\omega)$, have Riemann equivalent power spectra if for any $0<f_1<f_2<f_{\mathrm{s}}/2$
\begin{equation}
\int_{f_{1}}^{f_{2}}S_{r_{1}r_{1}}(j\omega)df=\int_{f_{1}}^{f_{2}}S_{r_{2}r_{2}}(j\omega)df+O(N^{-1})
\label{eq:PowerInBand}
\end{equation}
The $O(N^{-1})$ term, with $\frac{N}{2}-1$ the number of harmonics, is present if at least one of the signals is periodic. If $r_i(t)$ is periodic, then $S_{r_ir_i}(j\omega)$ is a sum of Dirac impulses and the integral in \eqref{eq:PowerInBand} is replaced by
\begin{equation}
\int_{f_{1}}^{f_{2}}S_{r_{i}r_{i}}(j\omega)df=\frac{1}{N}\sum_{k=k_{1}}^{k_{2}}\mathbb{E}\{|\hat{R}_{i}(\frac{k}{N}f_{\mathrm{s}})|^2\}\label{eq:PowerInBandPeriodic}
\end{equation}
with $\hat{R}_{i}(\frac{k}{N}f_{\mathrm{s}})/\sqrt{N}$ the  $k$-th Fourier coefficient, $k_1=\lceil N\frac{f_1}{f_{\mathrm{s}}}\rceil$, and $k_2=\lfloor N\frac{f_2}{f_{\mathrm{s}}}\rfloor$, where $\left\lceil x\right\rceil$ ($\left\lfloor x\right\rfloor$) is the smallest (largest) integer larger (smaller) than or equal to $x$. In addition, $S_{\hat{R}_i\hat{R}_i}(f)=\mathbb{E}\{|\hat{R_i}(f)|^2\}$ is a uniformly bounded function with a finite number of discontinuities on the interval $[0,f_{\mathrm{s}}/2]$.
\end{definition}
Using Definition \ref{def:RiemannEquivalentPowerSpectra}, the class of random phase multisines (Definition \ref{def:RandomPhaseMultisine}) and periodic noise (Definition \ref{def:PeriodicNoise}) signals can be extended to asymptotically ($N\rightarrow \infty$) normally distributed signals with Riemann equivalent power spectrum.
\begin{definition}[Class $\mathbb{U}$ of Asymptotically Normally Distributed Signals with Riemann Equivalent Power Spectrum]
\label{def:GaussianRiemannEquivalent}
$\mathbb{U}$ is the class of asymptotically ($N\rightarrow \infty$) normally distributed signals with Riemann equivalent power spectrum [see Definition \ref{def:RiemannEquivalentPowerSpectra}].
\end{definition}
Stationary Gaussian noise, random phase multisines (Definition \ref{def:RandomPhaseMultisine}), and periodic noise (Definition \ref{def:PeriodicNoise}) are examples of signals belonging to the Riemann equivalence class $\mathbb{U}$. The DC-value of the class $\mathbb{U}$ defines the set-point of the nonlinear system.
\section{Class of Nonlinear Systems}
\label{sec:ClassNonLinSystems}
In this section we consider the setup of Figure \ref{fig:EIV_NLTI_PN_ClosedLoop} without the input-output measurement noise sources $n_u(t)$ and $n_y(t)$. The resulting setup can be considered as a two-input $r(t)$ and $w(t)$, two-output $y(t)$ and $u(t)$ nonlinear system. Hence, to describe the class of nonlinear feedback systems for which the BLA framework is valid, we need the concept of a multiple-input, multiple-output finite degree Volterra series (see Section \ref{subsec:FiniteVolterraSeries}). Using this concept, the classes of nonlinear time-invariant systems without and with process noise $w(t)$ necessary to develop the BLA theory, are defined in Sections \ref{subeq:NLTI_no_PN} and \ref{subeq:NLTI_PN}, respectively.
\subsection{Finite Volterra Series}
\label{subsec:FiniteVolterraSeries}
\begin{definition}[Finite Degree Volterra Series]
\label{def:FiniteVolterraSeries}
The response $z(t)\in \mathbb{R}^{n_z}$ of a causal finite degree Volterra series to an input $x(t)\in \mathbb{R}^{n_x}$ has the form
\begin{subequations}
\label{subeq:FiniteVolterraSeries}
\begin{equation}
z(t)= \sum_{\alpha_1,\ldots,\alpha_{n_x}=0}^{K}z_{\alpha_1,\ldots,\alpha_{n_x}}(t)\label{eq:RespFiniteVolterraSeries}
\end{equation}
with $K\in\mathbb{N}$ the finite nonlinear degree. $z_{0,\ldots,0}(t)$ is a constant, and $z_{\alpha_1,\ldots,\alpha_{n_x}}(t)$ is defined through a multi-dimensional convolution integral of the kernel
\begin{equation}
 g_{\alpha_1,\ldots,\alpha_{n_x}}(\tau_{11},\ldots,\tau_{1\alpha_1},\ldots,\tau_{n_x1},\ldots,\tau_{n_x\alpha_{n_x}})\in \mathbb{R}^{n_z}
\label{eq:VectorKernel}
\end{equation}
and the $n_x$ input signals $x_{[l]}(t)$, $l=1,\ldots,n_x$,
\begin{align}
z_{\alpha_1,\ldots,\alpha_{n_x}}(t)=&\int_{0}^{\infty}\ldots\int_{0}^{\infty}g_{\alpha_1,\ldots,\alpha_{n_x}}(\tau_{11},\ldots,\tau_{n_x\alpha_{n_x}})\nonumber\\
&\prod_{l=1}^{n_x}\prod_{i=1}^{\alpha_l}x_{[l]}(t-\tau_{li})d\tau_{li}
\label{eq:ConvKernelDegreeAlpha}
\end{align}
If $\alpha_l=0$, then the product $\prod_{i=1}^{\alpha_l}\ldots$ in \eqref{eq:ConvKernelDegreeAlpha} is equal to one, and the kernel \eqref{eq:VectorKernel} does not depend on the corresponding $\tau_{li}$, $i=1,\ldots,\alpha_l$. 
\end{subequations}
\end{definition}
The kernel \eqref{eq:VectorKernel} can be interpreted as a multi-dimensional vector impulse response and is called the Volterra kernel of degree $\alpha=\sum_{l=1}^{n_x}\alpha_l$ \cite{Schetzen_book_2006}. The multi-dimensional integral \eqref{eq:ConvKernelDegreeAlpha} remains the same if the kernel is replaced by a symmetrized kernel which is the average of the original kernel over all $\prod_{l=1}^{n_x}\alpha_l!$ permutations within the $n_x$ groups of variables $\{\tau_{l1},\ldots,\tau_{l\alpha_l}\}$, $l=1,\ldots,n_x$.

The DFT \eqref{eq:DFT_sqrt_N_scaling} of the periodic steady state response $z(t)$ of the finite Volterra series \eqref{subeq:FiniteVolterraSeries} to $n_x$ random phase multisines or  periodic noise inputs $x(t)$ is given by
\begingroup
\allowdisplaybreaks
\begin{subequations}
\label{subeq:DFT_FiniteVolterraSeries}
\begin{align}
&Z(k) = \sum_{\alpha_1,\ldots,\alpha_{n_x}=0}^{K}Z_{\alpha_1,\ldots,\alpha_{n_x}}(k)~~\mathrm{for}~~k\ne 0\label{eq:DFT_PeriodicSteadyState_FiniteVolterraSeries}\\
&Z_{\alpha_1,\ldots,\alpha_{n_x}}(k) = \frac{1}{N^{\frac{\alpha-1}{2}}}\sum_{l=1}^{n_x}\sum_{i=1}^{\alpha_l}\sum_{k_{li}=-\frac{N}{2}+1}^{\frac{N}{2}-1}\nonumber\\
&G_{\alpha_1,\ldots,\alpha_{n_x}}(j\omega_{k_{11}},\ldots,j\omega_{k_{n_x\alpha_{n_x}}})\prod_{l=1}^{n_x}\prod_{i=1}^{\alpha_l}X_{[l]}(k_{li})\nonumber\\
&\mathrm{subject\,to}~k = \sum_{l=1}^{n_x}\sum_{i=1}^{\alpha_l}k_{li}~\mathrm{and\,with}~\alpha=\sum_{l=1}^{n_x}\alpha_l\label{eq:DFT_y_alpha}
\end{align}
\end{subequations}
\endgroup
[proof: see \cite{Chua_NG_ECS_1979}]. $G_{\alpha_1,\ldots,\alpha_{n_x}}(j\omega_{k_{11}},\ldots,j\omega_{k_{n_x\alpha_{n_x}}})$, with $\omega=2\pi f$, is the multi-dimensional Fourier transform of the symmetrized kernel \eqref{eq:VectorKernel} evaluated at the DFT frequencies $f_{k_{li}}=k_{li}f_{\mathrm{s}}/N$, $l=1,\ldots,n_x$ and $i=1,\ldots,\alpha_l$ \cite{Schetzen_book_2006}. Hence, the order of the angular frequencies in each group $\{\omega_{k_{l1}},\ldots,\omega_{k_{l\alpha_l}}\}$, $l=1,\ldots,n_x$, has no importance in \eqref{eq:DFT_y_alpha}.

\subsection{Nonlinear Systems without Process Noise}
\label{subeq:NLTI_no_PN}
Fading memory nonlinear systems excited by the class of Gaussian signals with the same Riemann equivalent power spectrum [see Definition \ref{def:GaussianRiemannEquivalent}] can be approximated arbitrarily well in mean squared sense by a finite degree Volterra series \eqref{subeq:FiniteVolterraSeries} [see \cite{Pintelon_book_2012,BoydChua_IEEECAS_1985}]. For this system class, the steady state response to a periodic excitation with period $T$, is periodic with the same period $T$. This excludes systems generating sub-harmonics, autonomous oscillations, bifurcations and chaos. However, hard nonlinearities such as clipping, dead zones, relays, quantizers, $\ldots$ are allowed. Although -- in general --  the Volterra series expansion of a nonlinear feedback system does not exist \cite{Schetzen_book_2006}, on a restricted input domain, the response of nonlinear feedback systems can be approximated arbitrarily well in mean squared sense by a finite degree Volterra series \eqref{subeq:FiniteVolterraSeries}. It motivates  the following definition of the class of nonlinear systems considered. 

\begin{definition}[Class $\mathbb{S_{\mathrm{NL}}}$ of Nonlinear Systems -- no Process Noise]
\label{def:NL_TI_systems}
Consider the setup of Figure \ref{fig:EIV_NLTI_PN_ClosedLoop}, where the measurement noise sources $n_u(t), n_y(t)$ and the process noise $w(t)$ are set to zero. $\mathbb{S_{\mathrm{NL}}}$ is the class of nonlinear time-invariant systems whose response $z(t) = [y(t)\,u(t)]^T$ to the input $x(t)=r(t)$, around the set-point $x_{\mathrm{DC}}=\mathbb{E}\{r(t)\}$ and $z_{\mathrm{DC}}=[\mathbb{E}\{y(t)\}\,\mathbb{E}\{u(t)\}]^T$, can be approximated arbitrarily well in mean squared sense by a stable one-input, two-output finite degree Volterra series \eqref{subeq:FiniteVolterraSeries} of sufficiently high nonlinear degree $K$, for the class $\mathbb{U}$ of asymptotically $(N\rightarrow \infty)$ normally distributed excitation signals $r(t)$ with Riemann equivalent power spectrum [see Definition \ref{def:GaussianRiemannEquivalent}]. In addition, there exists a positive definite matrix $C_1>0$ and a constant $C_2>0$ such that the DFT \eqref{eq:DFT_sqrt_N_scaling} of $z(t)$ \eqref{subeq:FiniteVolterraSeries} and $r(t)\in\mathbb{U}$ fulfill
\begingroup
\allowdisplaybreaks
\begin{subequations}
\label{subeq:ExistenceCrossAutoPower}
\begin{align}
&\lim_{K\rightarrow\infty}\mathbb{E}\{Z(k)Z^H(k)\} \leqslant C_{1}<\infty\label{eq:ExistenceAutoPower}\\
&\Bigl|\lim_{K\rightarrow\infty}\mathbb{E}\{Z(k)\overline{R(k)}\}\Bigr| \leqslant C_{2}<\infty\label{eq:ExistenceCrossPower}
\end{align}
\end{subequations}
\endgroup
for $k=1,2,\ldots,N/2-1$ and $N\rightarrow \infty$, and where the magnitude  in \eqref{eq:ExistenceCrossPower} is taken element-wise.
\end{definition}
Definition \ref{def:NL_TI_systems} guarantees the existence of the auto- and cross-power spectra (or spectral densities) of the reference $r(t)$ and the input-output signals $u(t)$ and $y(t)$. Conditions \eqref{subeq:ExistenceCrossAutoPower} also impose the convergence ($K\rightarrow\infty$) of the one-input, two-output finite degree Volterra series \eqref{subeq:FiniteVolterraSeries}, and the uniformly boundedness of the element-wise taken magnitudes of all $G_{\alpha_1,\ldots,\alpha_{n_x}}(j\omega_{k_{11}},\ldots,j\omega_{k_{n_x\alpha_{n_x}}})$ in \eqref{subeq:DFT_FiniteVolterraSeries}. Note that the stability of the closed loop system in Figure \ref{fig:EIV_NLTI_PN_ClosedLoop} without process noise is assured by the stability of open loop system from reference $r(t)$ to input-output $z(t) = [y(t)\,u(t)]^T$.

\subsection{Nonlinear Systems Subject to Process Noise}
\label{subeq:NLTI_PN}
To quantify the impact of the process noise $w(t)$ on the best linear approximation of the plant, the expected value -- conditioned on the reference signal $r(t)$ -- of the response of the nonlinear feedback system in Figure \ref{fig:EIV_NLTI_PN_ClosedLoop} is calculated. It requires a suitable assumption on the process noise $w(t)$. Note that a similar approach is utilized in \cite{AbdalmoatyHjalmarsson_AUT_2019} for estimating parametric nonlinear dynamical models of nonlinear systems subject to process noise.
\begin{assumption}[Process Noise]
\label{as:ProcessNoise}
The process noise $w(t)$ is a stationary Gaussian process with finite second order moments. It is independently distributed of the reference signal $r(t)$.
\end{assumption}

Under Assumption \ref{as:ProcessNoise}, the following important property of a finite Volterra series can be shown.
\begin{property}[Conditional Expected Value Finite Volterra Series]
\label{prop:ExpValueVolterra}
Consider the finite degree Volterra series \eqref{subeq:FiniteVolterraSeries} from input $x(t)=[r(t)\,w(t)]^T$ to output $z(t)=[y(t)\,u(t)]^T$. Under Assumption~\ref{as:ProcessNoise}, the system from input $x(t)=r(t)$ to the expected value of the output conditioned on $r(t)$, $\check{z}(t)=\mathbb{E}\{z(t)|r(t)\}$, defines a single-input, dual-output finite degree Volterra series \eqref{subeq:FiniteVolterraSeries} with kernels
\begin{align}
g_{\alpha_1}&(\tau_{11},\ldots,\tau_{1\alpha_1})=\nonumber\\
&\sum_{\alpha_2=0}^{K}\int_0^{\infty}\ldots\int_0^{\infty}g_{\alpha_1,\alpha_2}(\tau_{11},\ldots,\tau_{1\alpha_1},\tau_{21},\ldots,\tau_{2\alpha_2})\nonumber\\
&~~\mathbb{E}\{w(t-\tau_{21})\ldots w(t-\tau_{2\alpha_2})\}d\tau_{21}\ldots d\tau_{2\alpha_2}\label{eq:ExpValueKernel}
\end{align}
\end{property}
\proof{Direct application of $\mathbb{E}\{.|r(t)\}$ to \eqref{subeq:FiniteVolterraSeries} gives \eqref{eq:ExpValueKernel}. Under Assumption \ref{as:ProcessNoise}, the expected value in the right hand side of \eqref{eq:ExpValueKernel} can be written as the sum of products of finite second order moments \cite{Schetzen_book_2006} and, hence, is finite.}\\

Property \ref{prop:ExpValueVolterra} motivates the following definition of the class of nonlinear feedback systems subject to process noise.
\begin{definition}[Class $\mathbb{S}_{\mathrm{NL},w}$ of Nonlinear Systems subject to Process Noise]
\label{def:NL_TI_systems_PN}
Consider the setup of Figure \ref{fig:EIV_NLTI_PN_ClosedLoop}, where the measurement noise sources $n_u(t)$ and $n_y(t)$ are set to zero. $\mathbb{S}_{\mathrm{NL},w}$ is the class of nonlinear time-invariant systems whose response $z(t)=[y(t)\,u(t)]^T$ to the input $x(t)=[r(t)\,w(t)]^T$, around the set-point $x_{\mathrm{DC}}=[\mathbb{E}\{r(t)\}\,\mathbb{E}\{w(t)\}]^T$ and $z_{\mathrm{DC}}=[\mathbb{E}\{y(t)\}\,\mathbb{E}\{u(t)\}]^T$, can be approximated arbitrarily well in mean squared sense by a stable two-input, two-output finite degree Volterra series \eqref{subeq:FiniteVolterraSeries} of sufficiently high nonlinear degree $K$, for the signal class $\mathbb{U}$ [see Definition \ref{def:GaussianRiemannEquivalent}], and process noise $w(t)$ satisfying Assumption \ref{as:ProcessNoise}. In addition, $R(k)$ and $Z(k)$, the DFT \eqref{eq:DFT_sqrt_N_scaling} of, respectively, $r(t)\in\mathbb{U}$ and $z(t)$ \eqref{subeq:FiniteVolterraSeries} satisfy conditions \eqref{subeq:ExistenceCrossAutoPower}, where the expected values are taken w.r.t. $r(t)$ and $w(t)$.
\end{definition}
Definition \ref{def:NL_TI_systems_PN} guarantees the existence of the cross- and auto-power spectra (or spectral densities) of the reference $r(t)$, the input $u(t)$ and the output $y(t)$ signals in the presence of process noise $w(t)$. Conditions \eqref{subeq:ExistenceCrossAutoPower} impose the convergence ($K\rightarrow\infty$) of the two-input, two-output finite degree Volterra series \eqref{subeq:FiniteVolterraSeries} and its expected value w.r.t. the process noise, and the uniformly boundedness of the multi-dimensional Fourier transform of the kernels $g_{\alpha_1,\alpha_2}(\tau_{11},\ldots,\tau_{1\alpha_1},\tau_{21},\ldots,\tau_{2\alpha_2})$ and their expected value \eqref{eq:ExpValueKernel}.

Note that the stability of the closed loop system in Figure \ref{fig:EIV_NLTI_PN_ClosedLoop} is assured by the stability of the open loop system from reference $r(t)$ and process noise $w(t)$ to input-output $z(t) = [y(t)\,u(t)]^T$. Note also that the system class $\mathbb{S}_{\mathrm{NL},w}$ is a two-input, two-output version of the system class $\mathbb{S}_{\mathrm{NL}}$ [see Definition \ref{def:NL_TI_systems}]. The system class $\mathbb{S}_{\mathrm{NL},w}$ has the following key property.
\begin{lemma}[Property System Class $\mathbb{S}_{\mathrm{NL},w}$]
\label{lem:ExpValue_class_Sw}
The expected value w.r.t. the process noise $w(t)$ transforms the system class $\mathbb{S}_{\mathrm{NL},w}$ [see Definition \ref{def:NL_TI_systems_PN}] into the system class $\mathbb{S}_{\mathrm{NL}}$ [see Definition \ref{def:NL_TI_systems}].
\end{lemma}
\proof{see Appendix \ref{sec:ProofPropClassNL_PN}.}\\

\begin{figure}[t!] %[htbp]: here, top, bottom, page (follow by ! to enforce, e.g. b!)
\centering
\includegraphics[width=0.75\columnwidth]{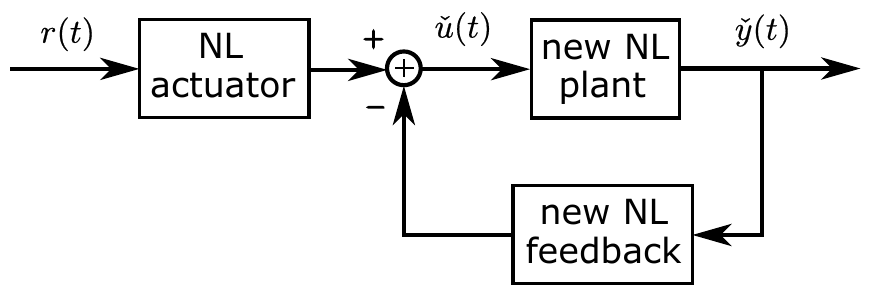} 
\caption{Taking the expected value conditioned on the reference signal $r(t)$ of the nonlinear feedback system $\in \mathbb{S}_{\mathrm{NL},w}$ [see Definition \ref{def:NL_TI_systems_PN} and Figure \ref{fig:EIV_NLTI_PN_ClosedLoop}], defines a new nonlinear feedback system $\in \mathbb{S}_{\mathrm{NL}}$ [see Definition \ref{def:NL_TI_systems}], with $\check{u}(t)=\mathbb{E}\{u(t)|r(t)\}$ and $\check{y}(t)=\mathbb{E}\{y(t)|r(t)\}$.}
\label{fig:EIV_NLTI_ExpValuePN_ClosedLoop}
\end{figure}

Lemma \ref{lem:ExpValue_class_Sw} motivates the block diagram shown in Figure \ref{fig:EIV_NLTI_ExpValuePN_ClosedLoop}, and justifies the definition of the best linear approximation given in Section \ref{sec:BestLinearApprox}.

\section{Best Linear Approximation}
\label{sec:BestLinearApprox}
First, assuming that no measurement noise is present, the best linear approximation (BLA) of nonlinear systems $\in\mathbb{S}_{\mathrm{NL},w}$ [see Definition \ref{def:NL_TI_systems_PN}] is defined and its properties are proven [Section \ref{subsec:BLA_DefProp}]. Next, the impact of the input-output measurement noise on the BLA framework is discussed [Section \ref{subsec:ImpactMeasNoise}]. Further, it is shown that the theory is also valid for discrete-time systems and the setup of Figure \ref{fig:EIV_NLTI_PN_ClosedLoop} is generalized to the case where the nonlinear actuator and feedback dynamics are also subject to process noise  [Section \ref{subsec:Gen_ActFB_PN}]. Finally, some nonparametric estimation methods are briefly discussed [Section \ref{subsec:NonparamEstim_BLA}] that allow one to detect and quantify the nonlinear behavior [Section \ref{subseq:Detect_NL}].

\subsection{Definition and Properties}
\label{subsec:BLA_DefProp}
Taking into account Lemma \ref{lem:ExpValue_class_Sw}, the BLA of nonlinear systems $\in \mathbb{S}_{\mathrm{NL},w}$ [see Definition \ref{def:NL_TI_systems_PN}] is defined as in \cite{Pintelon_etal_TIM_2013a} for nonlinear systems $\in \mathbb{S}_{\mathrm{NL}}$ [see Definition \ref{def:NL_TI_systems}]. The justification for the denotation `best' is given at the end of this subsection.
\begin{definition}[BLA in the Presence of Process Noise]
\label{def:BLA_PN}
The best linear approximation, $G_{\mathrm{BLA}}(j\omega)$, of a nonlinear system belonging to the class $\mathbb{S}_{\mathrm{NL},w}$ [see Definition \ref{def:NL_TI_systems_PN}] is defined as, for $k=1,2,\ldots\frac{N}{2}-1$,
\begingroup
\allowdisplaybreaks
\begin{subequations}
\label{subeq:G_BLA_PN}
\begin{align}
&G_{\mathrm{BLA},N}(j\omega_{k}) = \frac{\mathbb{E}\{Y(k)\overline{R(k)}\}}{\mathbb{E}\{U(k)\overline{R(k)}\}}\label{eq:G_BLA_PN_N}\\
&G_{\mathrm{BLA}}(j\omega) = \lim_{N\rightarrow\infty}G_{\mathrm{BLA},N}(j\omega_k)~~\mathrm{with}~~f_k=\frac{k}{N}f_{\mathrm{s}}\label{eq:G_BLA_PN}
\end{align}
\end{subequations}
\endgroup
where $\omega = 2\pi f$, with $f=\lim_{N\rightarrow\infty}f_k\in (0,0.5f_{\mathrm{s}})$, and where the expected values are taken w.r.t. the random realization of the reference $r(t)$ and the process noise $w(t)$.
\end{definition}
Using the BLA definition for nonlinear systems operating in open loop \cite{Pintelon_book_2012}, the BLA \eqref{eq:G_BLA_PN_N} of the nonlinear plant can be written as the ratio of the BLA $G_{RY,N}(j\omega_k)$ from reference to output and the BLA $G_{RU,N}(j\omega_k)$ from reference to input
\begingroup
\allowdisplaybreaks
\begin{subequations}
\label{subeq:BLAs_Ref_IO}
\begin{align}
G_{RY,N}(j\omega_k) &= \frac{\mathbb{E}\{Y(k)\overline{R(k)}\}}{\mathbb{E}\{|R(k)|^2\}}\label{eq:BLA_RY}\\
G_{RU,N}(j\omega_k) &= \frac{\mathbb{E}\{U(k)\overline{R(k)}\}}{\mathbb{E}\{|R(k)|^2\}}\label{eq:BLA_RU}
\end{align}
\end{subequations}
\endgroup
The difference between the actual response $z(t)=[y(t)\,u(t)]^T$ ($Z(k)$) of the nonlinear feedback system to the reference $r(t)$ ($R(k)$), and the response $z_{\mathrm{BLA}}(t)=[y_{\mathrm{BLA}}(t)\,u_{\mathrm{BLA}}(t)]^T$ ($Z_{\mathrm{BLA}}(k)$) predicted by the BLAs \eqref{subeq:BLAs_Ref_IO}, depends nonlinearly on the reference $r(t)$ and the process noise $w(t)$. It can be split into two different contributions:
\begin{enumerate}
\item The terms in $z(t)$ that do not depend on the actual realization of $w(t)$. Their sum is called the \textit{observed stochastic nonlinear distortion} $\tilde{z}_{\mathrm{S}}(t)$ ($\tilde{Z}_{\mathrm{S}}(k)$). $\tilde{Z}_{\mathrm{S}}(k)$ is formally defined as
\begin{equation}
\tilde{Z}_{\mathrm{S}}(k) = \mathbb{E}\{Z(k)|r(t)\} - Z_{\mathrm{BLA}}(k)
\label{eq:Def_Zs_PN}
\end{equation}
and it depends -- in general -- on the power spectral densities of $r(t)$ and $w(t)$. The latter is a major difference w.r.t. the classical framework without process noise.   
\item The terms that depend on the actual realization of $w(t)$. Their sum is called -- with some dual-use of terminology -- the \textit{observed process noise} $\tilde{z}_{\mathrm{P}}(t)$ ($\tilde{Z}_{\mathrm{P}}(k)$). $\tilde{Z}_{\mathrm{P}}(k)$ is formally defined as
\begin{align}
\tilde{Z}_{\mathrm{P}}(k) &= Z(k) - Z_{\mathrm{BLA}}(k) - \tilde{Z}_{\mathrm{S}}(k)\nonumber\\
 &= Z(k) - \mathbb{E}\{Z(k)|r(t)\}\label{eq:Def_Zp_PN}
\end{align}
Note that the observed process noise \eqref{eq:Def_Zp_PN} might depend on the actual realization of the reference $r(t)$. This is a major difference w.r.t. the linear case.
\end{enumerate}
It can easily be verified that the following condition holds
\begin{equation}
Z(k) = Z_{\mathrm{BLA}}(k)+\tilde{Z}_{\mathrm{S}}(k)+\tilde{Z}_{\mathrm{P}}(k)\label{eq:Zsplit}
\end{equation}
where
\begin{equation}
Z_{\mathrm{BLA}}(k)=\left[\begin{array}{c}
G_{RY,N}(j\omega_{k})\\
G_{RU,N}(j\omega_{k})
\end{array}\right]R(k)+\left[\begin{array}{c}
T_{G_{RY}}(j\omega_{k})\\
T_{G_{RU}}(j\omega_{k})
\end{array}\right]
\label{eq:ZBLA_R}
\end{equation}
and with $T_{G_{RY}}(j\omega_{k})$ and $T_{G_{RU}}(j\omega_{k})$ the transient terms \cite{Pintelon_book_2012}.

\begin{figure}[t!] %[htbp]: here, top, bottom, page (follow by ! to enforce, e.g. b!)
\centering
\includegraphics[width=0.8\columnwidth]{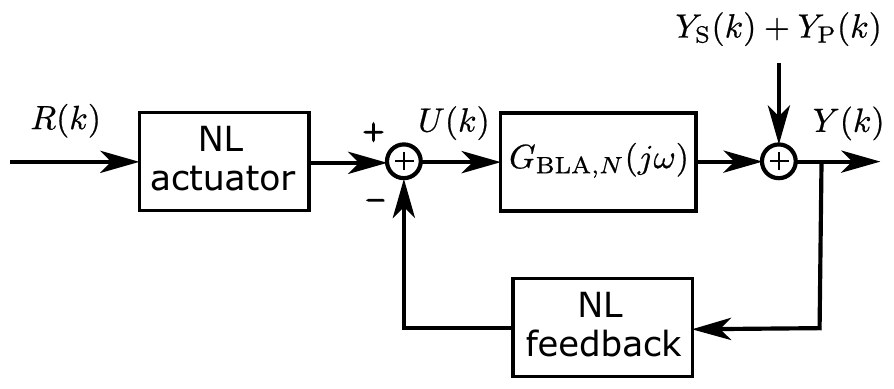} 
\caption{Best linear approximation $G_{\mathrm{BLA},N}(j\omega)$ \eqref{eq:G_BLA_PN_N} of a nonlinear system belonging to the class $\mathbb{S}_{\mathrm{NL},w}$ [see Definition \ref{def:NL_TI_systems_PN}]. The stochastic nonlinear distortion $Y_{\mathrm{S}}(k)$ and the process noise $Y_{\mathrm{P}}(k)$ are mutually uncorrelated and uncorrelated with -- but not independent of -- the reference $R(k)$.}
\label{fig:BLA_NLTI_PN_ClosedLoop}
\end{figure}

Using \eqref{eq:Zsplit}, \eqref{eq:ZBLA_R}, $G_{\mathrm{BLA},N}=G_{RY,N}/G_{RU,N}$ and $T_{G_{\mathrm{BLA}}}=T_{G_{RY}}/T_{G_{RU}}$, the difference between the actual output $Y(k)$ of the nonlinear plant and the output $Y_{\mathrm{BLA}}(k)$ predicted by the BLA \eqref{eq:G_BLA_PN_N} is readily found
\begin{subequations}
\label{subeq:Def_YS_YP}
\begin{align}
Y(k)-(G_{\mathrm{BLA},N}(j\omega_k)U(k)+&T_{G_{\mathrm{BLA}}}(j\omega_k)) =\nonumber\\
&Y_{\mathrm{S}}(k) + Y_{\mathrm{P}}(k)\label{eq:OutputResidual}
\end{align}
where $Y_{\mathrm{S}}(k)$ and $Y_{\mathrm{P}}(k)$ are, respectively, the stochastic nonlinear distortion and the process noise of the nonlinear plant 
\begin{align}
Y_{\mathrm{S}}(k) &= \tilde{Y}_{\mathrm{S}}(k) - G_{\mathrm{BLA},N}(j\omega_k)\tilde{U}_{\mathrm{S}}(k)\label{eq:Def_YS}\\
Y_{\mathrm{P}}(k) &= \tilde{Y}_{\mathrm{P}}(k) - G_{\mathrm{BLA},N}(j\omega_k)\tilde{U}_{\mathrm{P}}(k)\label{eq:Def_YP}
\end{align}
\end{subequations}
Figure \ref{fig:BLA_NLTI_PN_ClosedLoop} shows the corresponding block diagram.

The properties of the BLA \eqref{subeq:G_BLA_PN}, the stochastic nonlinear distortion \eqref{eq:Def_YS} and the process noise \eqref{eq:Def_YP} are established in the following theorem.

\begin{theorem}[Best Linear Approximation, Stochastic Nonlinear Distortion, and Process Noise]
\label{thm:Prop_BLA_YS_YP}
Consider the class of nonlinear systems $\mathbb{S}_{\mathrm{NL},w}$ [see Definition \ref{def:NL_TI_systems_PN}]. The best linear approximation \eqref{subeq:G_BLA_PN}, the stochastic nonlinear distortion \eqref{eq:Def_YS} and the process noise \eqref{eq:Def_YP} have the following properties for $k,l=1,2,\ldots,\frac{N}{2}-1$:
\begin{enumerate}
\item \label{enum:prop1_BLA}The BLA of the nonlinear plant from $u(t)$ to $y(t)$ (see Figure \ref{fig:EIV_NLTI_PN_ClosedLoop} -- no measurement noise) is equal to the BLA of the new nonlinear plant from $\mathbb{E}\{u(t)|r(t)\}$ to $\mathbb{E}\{y(t)|r(t)\}$ (see Figure \ref{fig:EIV_NLTI_ExpValuePN_ClosedLoop}).
\item \label{enum:prop2_BLA}$G_{\mathrm{BLA}}(j\omega)$ \eqref{eq:G_BLA_PN} is the same for all Gaussian-like signals $r(t)\in\mathbb{U}$ [see Definition \ref{def:GaussianRiemannEquivalent}], and only depends on the odd degree Volterra kernels \eqref{eq:ExpValueKernel} and the power spectral densities of $r(t)$ and $w(t)$. In addition \eqref{eq:G_BLA_PN_N} and \eqref{eq:G_BLA_PN} are related as
\begin{equation}
G_{\mathrm{BLA},N}(j\omega_k)=G_{\mathrm{BLA}}(j\omega_k)+O(N^{-1})
\label{eq:Rel_GblaN_Gbla}
\end{equation}
with $\omega_k=2\pi f_k$ and $f_k=k f_{\mathrm{s}}/N$.
\item \label{enum:prop_YS}$Y_{\mathrm{S}}(k)$ has the properties:
	\begin{enumerate}
	\item \label{subenum:YS_ZeroMean}Zero mean value
	\begin{subequations}
	\begin{equation}
	\mathbb{E}\{Y_{\mathrm{S}}(k)\}=0\label{eq:YS_ZeroMean}
	\end{equation}
	\item \label{subenum:YS_U_UnCorrelated}Asymptotically ($N\rightarrow\infty$) uncorrelated with -- but not independent of -- $R(k)$
	\begin{equation}
	\mathbb{E}\{Y_{\mathrm{S}}(k)\overline{R(k)}\} = O(N^{-1})\label{eq:YS_U_UnCorrelated}
	\end{equation}
	\item \label{subenum:YS_CircCompl}Asymptotically ($N\rightarrow\infty$) circular complex normally distributed
	\begin{equation}
	\mathbb{E}\{Y_{\mathrm{S}}^2(k)\} = O(N^{-1}) \label{eq:YS_CircCompl}
	\end{equation}
	\item \label{subenum:YS_UncorrFreq}Asymptotically ($N\rightarrow\infty$) uncorrelated over the frequencies
	\begin{equation}
	\mathbb{E}\{Y_{\mathrm{S}}(k)\overline{Y_{\mathrm{S}}(l)}\} = O(N^{-1})~~\mathrm{for}~~k\ne l \label{eq:YS_UncorrFreq}
	\end{equation}
	\end{subequations}
	\item \label{subenum:YS_SmoothVar}$\mathrm{var}(Y_{\mathrm{S}}(k))$ is a smooth function of the excited frequencies.
	\end{enumerate}
\item \label{enum:prop_YS_YP}$Y_{\mathrm{S}}(k)$ and $Y_{\mathrm{P}}(k)$ are mutually uncorrelated
\begin{equation}
\mathbb{E}\{Y_{\mathrm{S}}(k)\overline{Y_{\mathrm{P}}(l)}\} = 0~~\mathrm{and}~~\mathbb{E}\{Y_{\mathrm{S}}(k)Y_{\mathrm{P}}(l)\} = 0
\label{eq:prop_YS_YP}
\end{equation}
\item \label{enum:prop_YP}$Y_{\mathrm{P}}(k)$ has the same properties \ref{subenum:YS_ZeroMean}--\ref{subenum:YS_SmoothVar} as $Y_{\mathrm{S}}(k)$.
\end{enumerate}
\end{theorem}
\proof{See Appendix \ref{sec:ProofTheoremPropBLA_YS_YP}.\\}

In the remainder of this subsection we explain in which sense \eqref{subeq:G_BLA_PN} is the `best' approximation. Recall that \eqref{subeq:BLAs_Ref_IO} is the solution of the Wiener-Hopf equation
\begin{equation}
\mathrm{arg\,min}_{g(t)}\mathbb{E}\{\bigl\Vert z(t)-\mathbb{E}\{z(t)\}-g(t)\ast (r(t)-\mathbb{E}\{r(t)\})\bigr\Vert^2_2\}\nonumber
\end{equation}
with $z(t)=[y(t)\,u(t)]^T$, $g(t)$ the $2\times 1$ impulse response of the linear approximation, and $\ast$ the convolution product \cite{Eykhoff_book_1974,BendatPiersol_book_1980}. Hence, the spectral analysis estimate \eqref{subeq:BLAs_Ref_IO} is the `best' in the sense that it minimizes the mean squared difference between the zero mean part of the actual response and that of the linear approximation. This property is inherited by \eqref{subeq:G_BLA_PN} because it is the ratio of two best linear approximations \eqref{eq:BLA_RY} and \eqref{eq:BLA_RU}.

\subsection{Impact Measurement Noise}
\label{subsec:ImpactMeasNoise}
The impact of the measurement noise on the BLA \eqref{subeq:G_BLA_PN} is discussed under the following assumption.
\begin{assumption}[Measurement Noise]
\label{as:IO_MeasNoise}
The input $n_u(t)$ and output $n_y(t)$ measurement noise are -- possibly jointly correlated -- stationary random processes that are independent of the known reference signal $r(t)$ and the process noise $w(t)$. $n_u(t)$ and $n_y(t)$ have finite second order moments.
\end{assumption}
Under Assumption \ref{as:IO_MeasNoise}, the input $n_u(t)$ and output $n_y(t)$ measurement noise sources do not introduce a bias error in the expected values of \eqref{eq:G_BLA_PN_N}, which are now taken w.r.t. the random realizations of $r(t)$, $w(t)$, $n_u(t)$ and $n_y(t)$.

\subsection{Extensions}
\label{subsec:Gen_ActFB_PN}
The results of Theorem \ref{thm:Prop_BLA_YS_YP} are also valid for the discrete-time case because, at the sampling instants, a continuous-time Volterra system excited by a piecewise constant input can be described exactly by a discrete-time Volterra model [proof: see Appendix \ref{sec:StepInvTransfVolterra}]. Hence, for discrete-time systems, the impact of the process noise on the best linear approximation and the stochastic nonlinear distortion is exactly the same as for the continuous-time case.

\begin{figure}[t!] %[htbp]: here, top, bottom, page (follow by ! to enforce, e.g. b!)
\centering
\includegraphics[width=0.75\columnwidth]{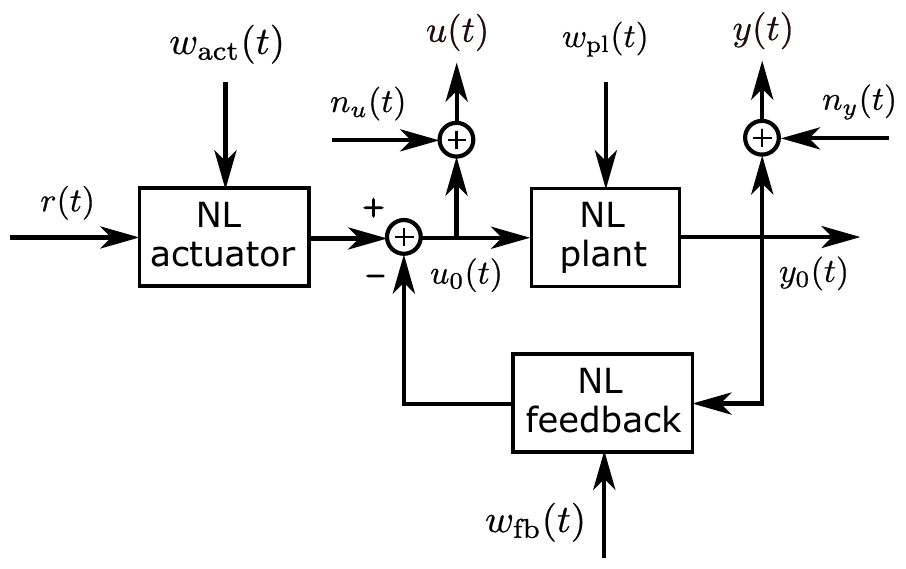} 
\caption{Noisy input $u(t)$, noisy output $y(t)$ measurement of a nonlinear time-invariant plant subject operating in closed loop. $w_{\mathrm{pl}}(t)$, $w_{\mathrm{act}}(t)$ and $w_{\mathrm{fb}}(t)$ are the process noise sources of, respectively, the plant, the actuator and the feedback. $n_u(t)$ and $n_y(t)$ are -- possibly jointly correlated -- stationary random processes that are independent of the known reference signal $r(t)$. The process noise sources are independently distributed of the reference and the input-output measurement noise.}
\label{fig:EIV_NLTI_fullPN_ClosedLoop}
\end{figure}

Consider the set-up shown in Figure \ref{fig:EIV_NLTI_fullPN_ClosedLoop}. If the process noise sources $w(t)=[w_{\mathrm{pl}}(t)\,w_{\mathrm{act}}(t)\,w_{\mathrm{fb}}(t)]^T$, satisfy a multivariate version of Assumption \ref{as:ProcessNoise}, and the system from input $x(t)=[r(t)\,w(t)]^T$ to output $z(t)=[y(t)\,u(t)]^T$ belongs to the system class $\mathbb{S}_{\mathrm{NL},w}$, then the results of Theorem \ref{thm:Prop_BLA_YS_YP} remain valid. Under Assumption \ref{as:IO_MeasNoise}, the measurement noise  does not affect the BLA \eqref{eq:G_BLA_PN_N}.

\subsection{Nonparametric Estimation}
\label{subsec:NonparamEstim_BLA}
There are basically two methods for estimating nonparametrically the best linear approximation, the variance of the nonlinear distortions and the noise variance due to the measurement and process noise. Both methods use random phase multisine signals $r(t)$ [see Definition \ref{def:RandomPhaseMultisine}] which belong to the class $\mathbb{U}$ of asymptotically ($N\rightarrow\infty$) normally distributed signals with Riemann equivalent power spectrum [see Definition \ref{def:RiemannEquivalentPowerSpectra}].

The \textit{robust method} [see \cite{Pintelon_book_2012}, p. 130] imposes no special conditions on the harmonic content of the random phase multisines \eqref{subeq:RandomPhase_MS}. All (odd) harmonics can be excited or some fraction can randomly be eliminated (random harmonic grid multisines). $P$ consecutive periods of the steady state response to a random phase multisine \eqref{subeq:RandomPhase_MS} are measured, and the  DFT \eqref{eq:DFT_sqrt_N_scaling} of each period of the known reference and the noisy input-output signals are calculated. This experiment is repeated for $M$ independent random phase realizations of the multisine with exactly the same harmonic content. Since the stochastic nonlinear distortion $y_{\mathrm{S}}(t)$ has the same periodicity of $r(t)$, the sample variances -- called \textit{noise variances} -- of the spectra over the $P$ consecutive periods only depend on the measurement and the process noise; while the sample variances -- called \textit{total variances} -- over the $M$ independent random phase realizations depend on the stochastic nonlinear distortion and the measurement and process noise. Subtracting the noise variances from the total quantifies the variance of the nonlinear distortions.

The \textit{fast method} [see \cite{Pintelon_book_2012}, p. 135] starts from one experiment with a full or odd random phase multisine \eqref{subeq:RandomPhase_MS} with random harmonic grid \cite{Schoukens_etal_TIM_2009}.  $P$ consecutive periods of the steady state response are measured, and the DFT \eqref{eq:DFT_sqrt_N_scaling} of each period of the known reference and the noisy input-output signals are calculated. At the non-excited frequencies of the random phase multisine $r(t)$, the input-output spectra only depend on the stochastic nonlinear distortion, the process noise and the measurement noise and, hence, their magnitudes quantify the total standard deviation. Comparison of the total standard deviation with the sample standard deviation over the periods (= noise standard deviation) quantifies the level of the nonlinear distortions.

Transients due to the plant and/or disturbing noise dynamics increase the variability of the robust and fast BLA estimates and introduce a bias error (plant transients only). Therefore, using the smoothness of the BLA and the plant and noise transient terms as a function of the frequency,  robustified versions of the fast and robust methods have been developed  that decrease significantly the impact of the transients on the estimates. These methods are based on a \textit{local polynomial} [see \cite{Pintelon_book_2012}, Chapter 7] or a \textit{local rational} \cite{Peumans_etal_TIM_2018} approximation of the BLA and the transient terms.

\subsection{Detection of the Nonlinear Behavior}
\label{subseq:Detect_NL}
Using the nonparametric techniques of Section \ref{subsec:NonparamEstim_BLA}, we can distinguish two types of nonlinear contributions of the plant dynamics:
\begin{itemize}
\item Type I: Nonlinear relationship between the input $u(t)$ and the output $y(t)$ that is independent of the actual realization of the process noise $w(t)$. This part of the response only affects the BLA and/or the nonlinear distortion $y_{\mathrm{S}}(t)$.
\item Type II: Nonlinear interactions between the input $u(t)$ and the process noise $w(t)$ that only influence the process noise $y_{\mathrm{P}}(t)$.
\end{itemize}
Note that a particular nonlinear term can contribute to both types of nonlinearities. Consider, for example, a system operating in open loop with the nonlinear term $u^2(t)w^2(t)$. Using definitions \eqref{eq:Def_Zs_PN} and \eqref{eq:Def_Zp_PN}, where $r(t)$ is replaced by $u(t)$, the term can be split as
\begin{equation}
u^2(t)w^2(t) = u^2(t)(w^2(t)-\sigma^2_w(t)) + (u^2(t)-\gamma_u)\sigma^2_w+\gamma_u\sigma^2_w\nonumber
\end{equation}
with $\gamma_u=\mathbb{E}\{u^2(t)\}$, and where the first and the second term in the right hand side only contribute to, respectively, $y_{\mathrm{P}}(t)$ (Type II nonlinearity) and $y_{\mathrm{S}}(t)$ (Type I nonlinearity). 

If the total variance is larger than the noise variance, then Type I and/or Type II nonlinearities are present. On the other hand, if the total variance is equal to the noise variance, then it is likely that the system behaves linearly. However, the Type I and Type II nonlinearities can be hidden by the measurement noise and/or the process noise [see Sections \ref{sec:SimulationExample} and \ref{sec:MeasurementExample}]. 

To distinguish the Type I from the Type II contributions (total variance $>$ noise variance) and/or to confirm or reject the hypothesis of a linear system (total variance $=$ noise variance), additional experiments with one or more different reference power spectral densities are required. Based on the (lack of) variation of the BLA, the variance of the nonlinear distorions and the noise variance, the presence of the Type I and Type II nonlinear contributions can be detected as shown in Table \ref{tab:ConnTypeI_II_BLA_NL_noise}.

\begin{table} [h!]
\centering
\begin{threeparttable}
\caption{Connection between the Type I and Type II nonlinearities and the impact of the reference power on the BLA, its noise variance $\sigma_{\mathrm{BLA,n}}^2$ and its variance due to the nonlinear distortion $\sigma_{\mathrm{BLA,S}}^2$.}
{
% \begin{tabular}{p{0.1cm}ccp{0.1cm}}
\begin{tabular}{|l|l|l|}
%\toprule 
\multicolumn{3}{l}{Change in reference power results in:}
 \\
\hline
%\vspace{0.1mm}
 & & \\
 BLA changes? &$\sigma_{\mathrm{BLA,S}}^2$ changes?  &$\sigma_{\mathrm{BLA,n}}^2$ inversely proportional\\
   & &to the reference power?\\

%\midrule
%\vspace{0.1mm}\\
%\hline
%\vspace{0.1mm}\\
 & & \\
yes $\rightarrow$ Type I &yes $\rightarrow$ Type I &yes $\rightarrow$ not Type II\\
no  $\rightarrow$ undecided\tnote{1} &no $\rightarrow$ not Type I\tnote{2} &no $\rightarrow$ Type II\\
%\botrule
%\vspace{0.1mm}\\
& & \\
\hline
\end{tabular}
}
\begin{tablenotes}
\item[1] even degree nonlinearities do not contribute to the BLA, \\
$~~~$but do contribute to $\sigma_{\mathrm{BLA,S}}^2$\\
\item[2] if the BLA does not change
\end{tablenotes}
{% possibility to add a footnote
}
\label{tab:ConnTypeI_II_BLA_NL_noise}
\end{threeparttable}
\end{table}
\section{Simulation Example}
\label{sec:SimulationExample}
Analytic calculation of the impact of the process noise on the BLA of a nonlinear system operating in feedback is possible for the following nonlinear finite impulse response (NFIR) system
\begin{subequations}
\label{subeq:Feedback_NFIR_PN}
\begin{align}
y(t) &= u(t-1) + u(t-2)w^2(t)\label{eq:NFIR_IO}\\
u(t) &= r(t) - \alpha y(t)\label{eq:Feedback_NFIR}
\end{align}
\end{subequations}
with $t\in\mathbb{Z}$. The reference signal $r(t)$ is a zero mean random phase multisine \eqref{subeq:RandomPhase_MS}, with $N=1024$ and $R_0=0$. All amplitudes $|R_k|$, $k=\pm 1,\pm 2,\ldots,\pm N/2-1$ are equal and chosen such that the standard deviation of $r(t)$ is equal to one ($\sigma_r=\mathrm{std}(r(t))=1$). The process noise $w(t)$ is zero mean discrete-time white Gaussian noise with variance $\mathrm{var}(w(t))$. For stability reasons $\alpha$ in \eqref{eq:Feedback_NFIR} is constrained as
\begin{subequations}
\label{subeq:Constraints_alpha}
\begin{align}
&0<\alpha<\mathrm{min}(4\sigma^2_w,\sigma^{-2}_w)&\mathrm{for}~~\sigma_w\ne0\label{eq:Constraint_alpha_PN}\\
&|\alpha|<1&\mathrm{for}~~\sigma_w=0\label{eq:Constraint_alpha_zero_PN}
\end{align}
\end{subequations}
[proof: see Appendix \ref{sec:Proof_BLA_Feedback_NFIR_PN}]. Here, the choice $\alpha=0.3$ is made.

The true values of the best linear approximation and its total variance, the process noise, and the stochastic nonlinear distortion, equal
\begingroup
\allowdisplaybreaks
\begin{subequations}
\label{subeq:BLA_feedback_NFIR_PN}
\begin{align}
&G_{\mathrm{BLA}}(j\omega) = e^{-j\omega T_{\mathrm{s}}} + \sigma_w^2 e^{-2j\omega T_{\mathrm{s}}}\label{eq:BLA_feedback_NFIR_PN}\\
&\mathrm{var}(\hat{G}_{\mathrm{BLA}}(j\omega)) \approx |1+\alpha G_{\mathrm{BLA}}(j\omega)|^2\frac{2\sigma_u^2\sigma_w^4}{\sigma_r^2}\label{eq:varBLA_feedback_NFIR_PN}\\
&y_{\mathrm{S}}(t) = 0\label{eq:ys_feedback_NFIR_PN}\\
&y_{\mathrm{P}}(t) = u(t-2)[w^2(t)-\sigma_w^2]\label{eq:yp_feedback_NFIR_PN}
\end{align}
\end{subequations}
\endgroup
with $\sigma_u^2=\mathrm{var}(u(t))$, and where $\sigma_u^2/\sigma_r^2$ is independent of $\sigma_r^2$ [proof: see Appendix \ref{sec:Proof_BLA_Feedback_NFIR_PN}]. Hence, the BLA and its total variance are independent of the variance of the reference signal. While the property of the BLA is consistent with an LTI system, that of the variance is not. The latter is due to the nonlinear interaction between the input and the process noise.

Starting from $P=2$ consecutive periods of the transient response to the random phase multisine $r(t)$, the fast local polynomial estimates of the best linear approximation and its total and noise variances are calculated from the known reference $r(t)$ and the noisy input $u(t)$ -- output $y(t)$ signals [see \cite{Pintelon_book_2012}, Chapter 7, for the details]. For this purpose, a second order local polynomial approximation ($R=2$) of the transient and the best linear approximation with ten degrees of freedom ($\mathrm{dof}=10$) is used. Given their high variability, the estimates are averaged over $M=100$ independent realizations of $r(t)$ and $w(t)$.
\begin{figure}[t!] %[htbp]: here, top, bottom, page (follow by ! to enforce, e.g. b!)
\centering
\includegraphics[width=0.55\columnwidth]{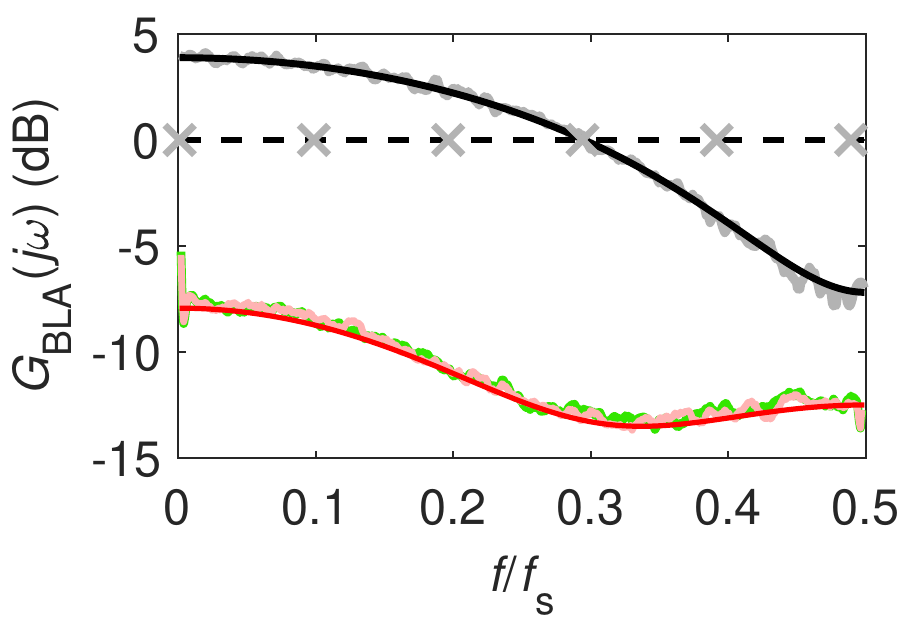} 
\caption{Best linear approximation \eqref{subeq:G_BLA_PN} of the closed loop NFIR system \eqref{subeq:Feedback_NFIR_PN}, with $\alpha=0.3$, for $\sigma_w=0$ [black dashes: true value; gray `$\times$': estimate] and $\sigma_w=0.75$ [black: true value, gray: estimate]. Fast estimates averaged over $M=100$ independent realizations of $r(t)$ and $w(t)$ for the case $\sigma_w=0.75$: the BLA [gray: estimate, black: true value], its total variance [pink: estimate, red: true value] and noise variance [green: estimate].}
\label{fig:BLA_Feedback_NFIR_PN}
\end{figure}

Figure \ref{fig:BLA_Feedback_NFIR_PN} shows the results for the cases $\sigma_w = 0$ and $\sigma_w = 0.75$. It can be seen that the estimates of the BLA and its total variance coincide with the true values \eqref{eq:BLA_feedback_NFIR_PN} and \eqref{eq:varBLA_feedback_NFIR_PN} divided by $\mathrm{dof}$ [the local polynomial approximation of the BLA reduces the variance of the estimate by a factor $\mathrm{dof}$]. Note that in the absence of process noise, $\sigma_w=0$, the feedback system \eqref{subeq:Feedback_NFIR_PN} is linear and noiseless, which results in a BLA estimate with zero variability. Note also that the shape of the BLA strongly depends on $\sigma_w^2$ [compare the black dashes with the black line].

Despite the nonlinear interaction between the input and the process noise, it follows from Figure \ref{fig:BLA_Feedback_NFIR_PN} that the total (red) and noise (green) variances of the BLA estimate coincide. It illustrates that -- similar to the measurement noise -- the process noise can hide the nonlinear behavior in FRF estimates. To reveal the nonlinear behavior, the BLA and its variance should be calculated for (two) different values of $\sigma_r^2$. In this simulation example, changing $\sigma_r^2$ will not modify the BLA \eqref{eq:BLA_feedback_NFIR_PN} nor its total variance \eqref{eq:varBLA_feedback_NFIR_PN}. The first observation is due to the linear input-output relation \eqref{eq:NFIR_IO}, while the second observation originates from the nonlinear process noise-input interaction in \eqref{eq:NFIR_IO}. 

\section{Measurement Example}
\label{sec:MeasurementExample}
\begin{figure}[t!] %[htbp]: here, top, bottom, page (follow by ! to enforce, e.g. b!)
\centering
\includegraphics[width=0.9\columnwidth]{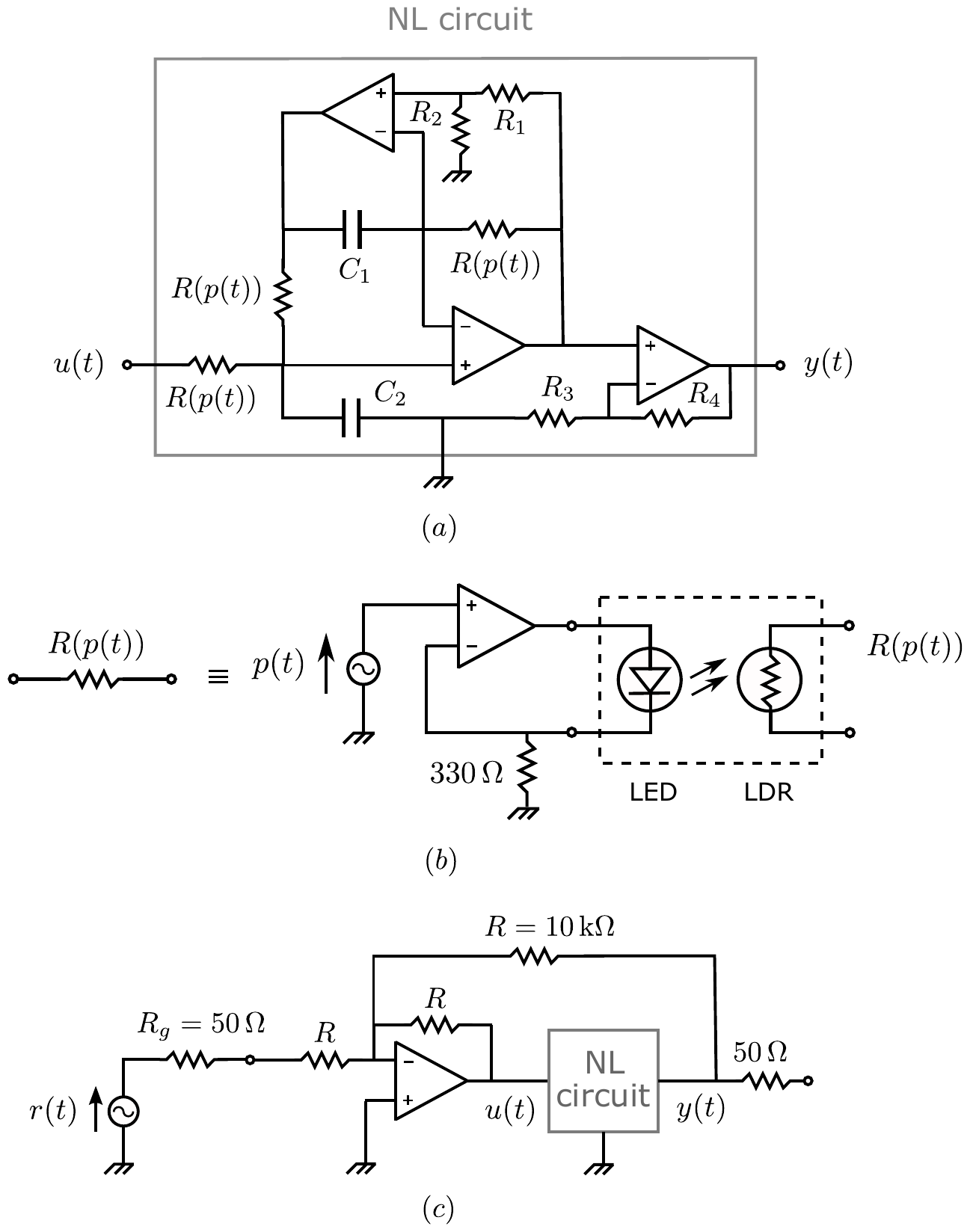} 
\caption{Nonlinear electrical circuit (a) operating in closed loop (c). It consists of three high gain operational ammplifiers (TL071), three voltage-dependent resistors $R(p(t))$, four resistors ($R_1=R_2=10\,\textrm{k}\Omega$, $R_3=5.31\,\textrm{k}\Omega$, and $R_4=100.8\,\textrm{k}\Omega$), and two capacitors ($C_1=C_2=10\,\textrm{nF}$). Schematic (b) shows the practical realization of $R(p(t))$. It is made using an operational amplifier, a $330\,\Omega$ resistor, and an electro-optical component (VTL5C1) consisting of a light-dependent resistor (LDR) and a light-emitting diode (LED). The voltage $p(t)=p_0+w(t)$, with $p_0$ the DC-value and $w(t)$ the process noise.}
\label{fig:NonlinearFeedback_ProcessNoise}
\end{figure}
Three experiments are performed on a nonlinear electronic circuit operating in feedback [see Figure \ref{fig:NonlinearFeedback_ProcessNoise}(c)]. The electronic circuit [see Figure \ref{fig:NonlinearFeedback_ProcessNoise}(a,b)] is a high gain bandpass filter whose nonlinear behavior is due to the nonlinearity of the operational amplifier and voltage-dependent resistor characteristics. The process noise $w(t)$ is introduced in the circuit via the voltage $p(t)$ of the voltage-dependent resistors $R(p(t))$
\begin{equation}
p(t) = p_0 + w(t)
\label{eq:Voltage_Resistors}
\end{equation}
where $p_0=1.6\,\textrm{V}$. At the sampling instances, $w(t)$ is a zero mean, white Gaussian noise process with standard deviation $\sigma_w$.

\begin{figure}[t!] %[htbp]: here, top, bottom, page (follow by ! to enforce, e.g. b!)
\centering
\includegraphics[width=0.49\columnwidth]{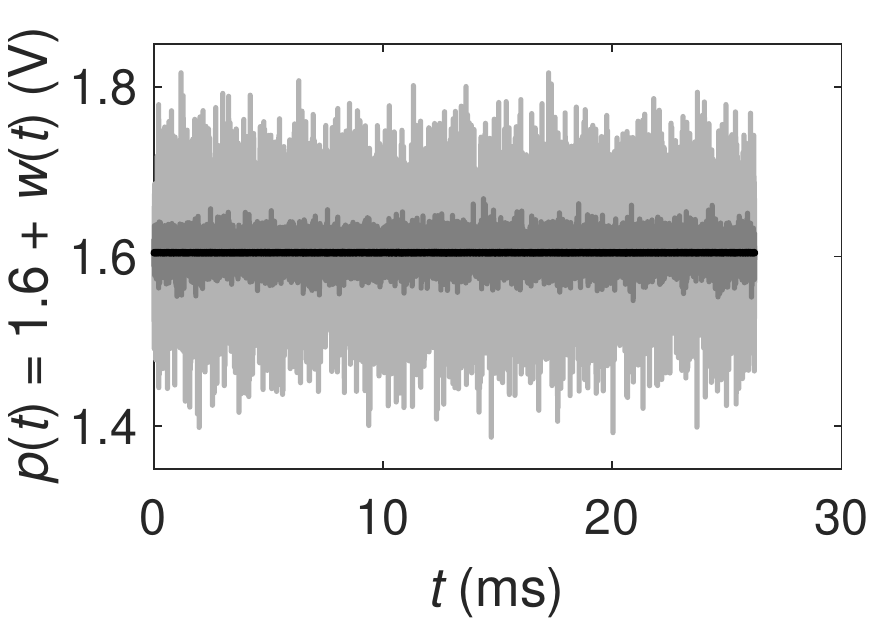}\\ 
\includegraphics[width=0.49\columnwidth]{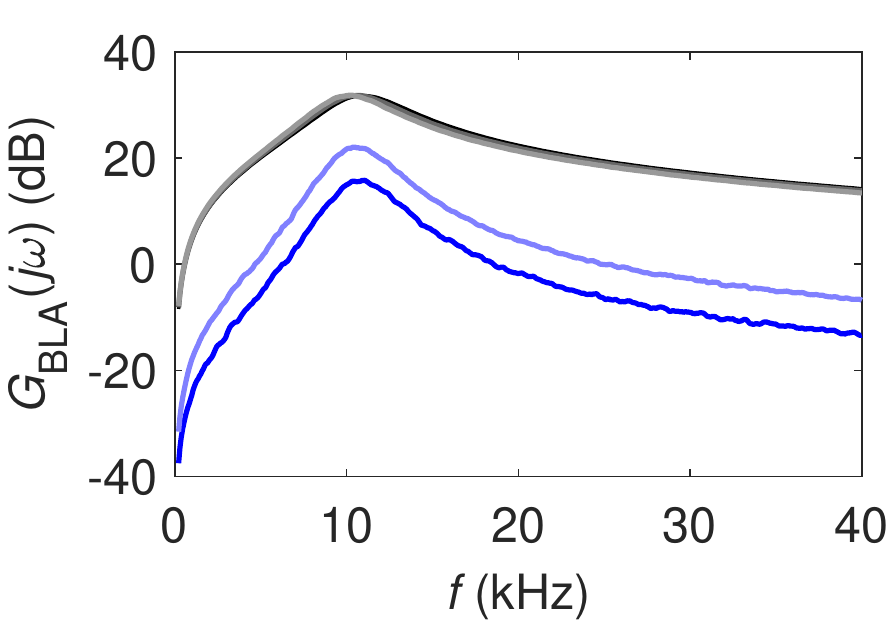}
\includegraphics[width=0.49\columnwidth]{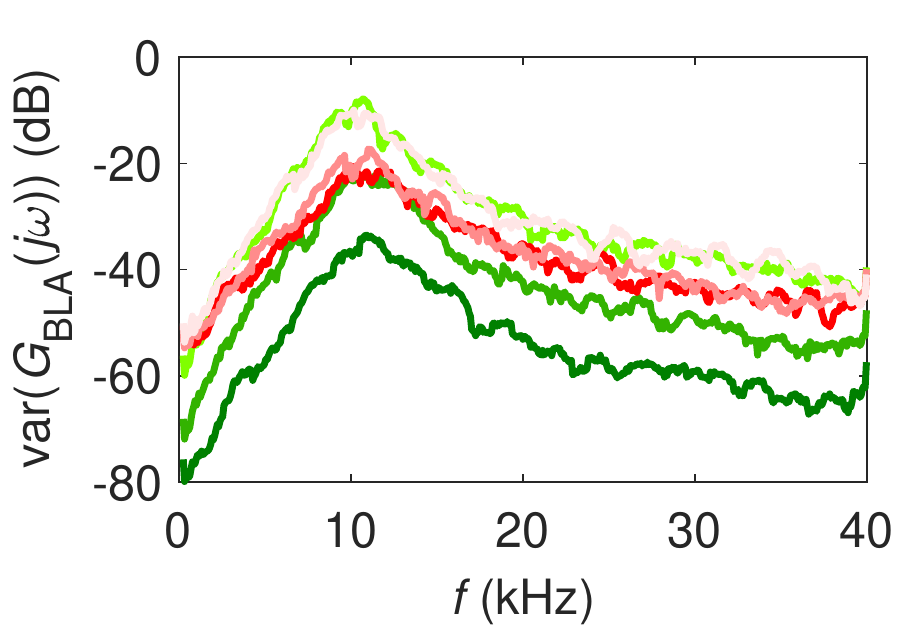} 
\caption{Best linear approximation (BLA) of the nonlinear circuit in Figure \ref{fig:NonlinearFeedback_ProcessNoise}(a) for three different values of the process noise $w(t)$ standard deviation $\sigma_w$. Top: voltage $p(t)$; bottom left: BLAs [black/gray lines] and the magnitude of the complex difference w.r.t. the zero process noise BLA [blue lines]; bottom right: noise [green lines] and total [red lines] variances of the BLAs. Black, dark green and red: $\sigma_w=0$; dark gray, dark blue, medium green and dark pink: $\sigma_w=14.8\,\textrm{mV}$; light gray, light blue and light pink: $\sigma_w=58.2\,\textrm{mV}$.}
\label{fig:BLA_NonlinearFeedback_ProcessNoise}
\end{figure}

For each experiment, the reference signal $r(t)$ is a zero mean random phase multisine \eqref{def:RandomPhaseMultisine} consisting of the sum of 522 sinewaves with uniformly distributed phases $\angle{R_k}$ and equal amplitudes $|R_k|$ in the band $[228.9\,\textrm{Hz},\,39.98\,\textrm{kHz}]$ chosen such that the standard deviation of $r(t)$ equals $1.34\,\textrm{V}$ ($f_{\mathrm{s}}=625\,\textrm{kHz}$, $N=16384$, $|R_{\pm k}|=A$ for $k=3,4,\ldots,524$ and $|R_{\pm k}|=0$ for $k=0,1,2,525,526$, $\ldots$, $N/2-1$). $P=2$ consecutive periods of the transient response of the input $u(t)$ and output $y(t)$ are acquired using a band-limited measurement setup (all signals are lowpass filtered before sampling). In the first experiment the process noise in \eqref{eq:Voltage_Resistors} is set to zero [$\sigma_w=0$], while in the second and third experiments $\sigma_w=14.8\,\textrm{mV}$ and $\sigma_w=58.2\,\textrm{mV}$, respectively [see Figure \ref{fig:BLA_NonlinearFeedback_ProcessNoise}, top]. The linear resistors and operational amplifiers also add some process noise to the circuit but in the second and third experiments their contribution can be neglected w.r.t. the externally applied $w(t)$. 

Via a fourth order local polynomial approximation over twelve neighbouring non-excited frequencies of the transient, and a fourth order local polynomial approximation over eleven neighboring excited frequencies of the frequency response function, the BLA and its noise and total variances are estimated from the known reference $r(t)$ and the noisy input $u(t)$ and output $y(t)$ signals [see \cite{Pintelon_book_2012}, Chapter 7, for the details]. Figure \ref{fig:BLA_NonlinearFeedback_ProcessNoise} shows the results.

The bottom left plot of Figure \ref{fig:BLA_NonlinearFeedback_ProcessNoise} shows the impact of the process noise on the BLA: the resonance shifts to the left for increasing values of $\sigma_w$. This can only be explained by a nonlinear interaction between the process noise $w(t)$ and the input $u(t)$. Indeed, the differences between the BLAs (blue lines) are well above the total variances of the BLA estimates (red/pink lines). From the bottom right plot it can be seen that the total variance (red and pink lines) is well above the noise variance (dark and medium green lines) for the first two experiments ($\sigma_w=0$ and $\sigma_w=14.8\,\textrm{mV}$), which reveals the nonlinear behavior of the electrical circuit. However, for the third experiment ($\sigma_w=58.2\,\textrm{mV}$), the total variance (light pink line) coincides with the noise variance (light green line). Due to the increased variability of the BLA estimate -- caused by the process noise -- the nonlinear behavior is hidden.

\section{Conclusion}
\label{sec:Conclusions}
The properties of the best linear approximation (BLA) of a certain class of continuous-time nonlinear feedback systems subject to process noise have been studied in detail. Compared with the open loop case \cite{Schoukens_etal_TAC_2020}, the BLA $G_{\mathrm{BLA}}(j\omega)$, the stochastic nonlinear distortion $Y_{\mathrm{S}}(k)$ and the process noise $Y_{\mathrm{P}}(k)$ depend on the reference $r(t)$ instead of the input $u(t)$. Compared with the case without process noise $w(t)$, $G_{\mathrm{BLA}}(j\omega)$ and $Y_{\mathrm{S}}(k)$ depend on the power spectral density of $w(t)$. By construction, $Y_{\mathrm{S}}(k)$ does not depend on the actual realization of $w(t)$, while $Y_{\mathrm{P}}(k)$ does. $Y_{\mathrm{S}}(k)$ and -- in general -- $Y_{\mathrm{P}}(k)$ depend on the actual realization of $r(t)$.

Using random phase multisine excitations, it is possible to estimate nonparametrically the BLA, the variance of the nonlinear distortions, and the noise variance due to the process and input-output measurement noise. Similar to the measurement noise, the process noise can mask the nonlinear behavior (total variance = noise variance); even in the case of a nonlinear interaction between the process noise $w(t)$ and the input $u(t)$. Unlike the measurement noise, the process noise power spectral density affects the BLA. Finally, a multiple experiment procedure is proposed to confirm or reject the linearity hypothesis (total variance = noise variance), and/or to distinguish nonlinear input-output behavior from nonlinear input-process noise interactions (total variance $>$ noise variance).

\appendices
\section{Proof of Equation \eqref{eq:AsVar_MS}}
\label{sec:ProofAsVarMultiSine}
The variance of $r(t)$ \eqref{eq:RandomPhase_MS} is given by
\begin{equation}
\mathrm{var}(r(t)) =\sum_{k,\,l=-\frac{N}{2}+1,\,k,l\ne 0}^{\frac{N}{2}-1}\mathbb{E}\left\{ R_{k}R_{l}\right\} e^{j2\pi\frac{k+l}{N}f_{\mathrm{s}}t}
\label{eq:Var_u}
\end{equation}
Since the amplitudes and phases of the Fourier coefficients are -- by construction -- independently distributed, we find for $\mathbb{E}\left\{ R_{k}R_{l}\right\}$
\begin{equation}
\mathbb{E}\left\{ R_{k}R_{l}\right\} =\begin{cases}
0 & k\ne \pm l\\
\mathbb{E}\{\left|R_{k}\right|^{2}\} & k=-l\\
\mathbb{E}\{\left|R_{k}\right|^{2}\}\mathbb{E}\left\{e^{j2\angle R_{k}}\right\}  & k=l
\end{cases}\label{eq:Mean_UkUl}
\end{equation}
Combining \eqref{eq:RandomPhaseConditions} with \eqref{eq:Mean_UkUl}, allows us to simplify \eqref{eq:Var_u} as
\begin{equation}
\mathbb{E}\left\{ r^{2}(t)\right\} =\frac{2}{N}\sum_{k=1}^{\frac{N}{2}-1}\mathbb{E}\Bigl\{\bigl|\hat{R}(\frac{k}{N}f_{\mathrm{s}})\bigr|^2\Bigr\}
\label{eq:RiemannSumVar}
\end{equation}
Taking the limit for $N\rightarrow \infty$ of the Riemann sum \eqref{eq:RiemannSumVar} finally proves \eqref{eq:AsVar_MS}.
\section{Proof of Lemma \ref{lem:ExpValue_class_Sw}}
\label{sec:ProofPropClassNL_PN}
To prove the lemma, we will show that conditions \eqref{subeq:ExistenceCrossAutoPower}, where the expected values are taken w.r.t. $r(t)$ and $w(t)$, imply
\begingroup
\allowdisplaybreaks
\begin{subequations}
\label{subeq:ExistenceCrossAutoPower_PN}
\begin{align}
&\lim_{K\rightarrow\infty}\mathbb{E}\{\check{Z}(k)\check{Z}^H(k)\} \leqslant C_{1}<\infty\label{eq:ExistenceAutoPower_PN}\\
&\Bigl|\lim_{K\rightarrow\infty}\mathbb{E}\{\check{Z}(k)\overline{R(k)}\}\Bigr| \leqslant C_{2}<\infty\label{eq:ExistenceCrossPower_PN}
\end{align}
\end{subequations}
\endgroup
with $\check{Z}(k)=\mathbb{E}\{Z(k)|r(t)\}$.

Since $u(t)$ and $w(t)$ are independently distributed [Assumption \ref{as:ProcessNoise}], the expected values in \eqref{subeq:ExistenceCrossAutoPower} can be calculated as
\begin{equation}
\mathbb{E}\{.\}=\mathbb{E}\{\mathbb{E}\{.|r(t)\}\}
\label{eq:SplitExpValue}
\end{equation}
Applying \eqref{eq:SplitExpValue} to \eqref{eq:ExistenceAutoPower}, taking into account that $\mathbb{E}\{ZZ^H\}=\mathrm{Cov}(Z)+\mathbb{E}\{Z\}\mathbb{E}\{Z\}^H \geqslant \mathbb{E}\{Z\}\mathbb{E}\{Z\}^H$, we find
\begin{equation}
\mathbb{E}\{Z(k)Z^H(k)\}\geqslant \mathbb{E}\{\check{Z}(k)\check{Z}^H(k)\}
\label{eq:ExpValueZZH}
\end{equation}
Combining \eqref{eq:SplitExpValue} with \eqref{eq:ExistenceCrossPower}, gives
\begin{equation}
\mathbb{E}\{Z(k)\overline{R(k)}\}=\mathbb{E}\{\check{Z}(k)\overline{R(k)}\}
\label{eq:ExpValueZconjR}
\end{equation}
Collecting \eqref{subeq:ExistenceCrossAutoPower}, \eqref{eq:ExpValueZZH} and \eqref{eq:ExpValueZconjR} proves \eqref{subeq:ExistenceCrossAutoPower_PN}.

\section{Proof of Theorem \ref{thm:Prop_BLA_YS_YP}}
\label{sec:ProofTheoremPropBLA_YS_YP}
Since $r(t)$ and $w(t)$ are independently distributed [Assumption \ref{as:ProcessNoise}], the expected values in \eqref{eq:G_BLA_PN_N} can be calculated as in \eqref{eq:SplitExpValue}, for example,
\begin{align}
\mathbb{E}\{Y(k)\overline{R(k)}\}&=\mathbb{E}\{\mathbb{E}\{Y(k)\overline{R(k)}\}|r(t)\}\nonumber\\
&=\mathbb{E}\{\mathbb{E}\{Y(k)|r(t)\}\overline{R(k)}\}\nonumber
\end{align}
and similarly for $\mathbb{E}\{U(k)\overline{R(k)}\}$, which proves Property \ref{enum:prop1_BLA} of the theorem.

\begin{figure}[t!] %[htbp]: here, top, bottom, page (follow by ! to enforce, e.g. b!)
\centering
\includegraphics[width=0.8\columnwidth]{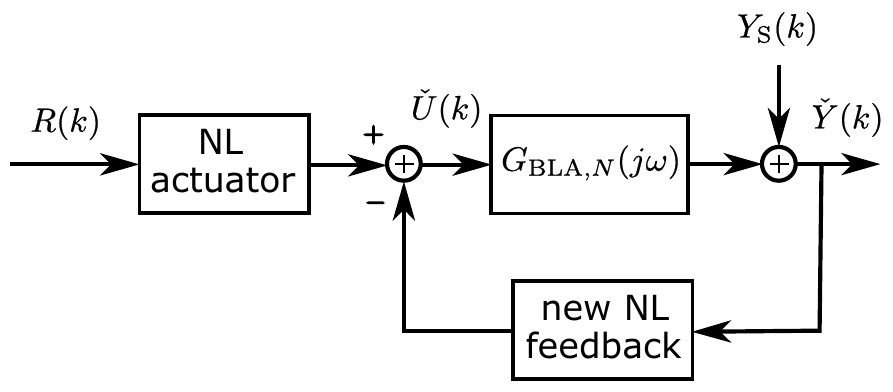} 
\caption{Best linear approximation $G_{\mathrm{BLA},N}(j\omega)$ \eqref{eq:G_BLA_PN_N} of a nonlinear system belonging to the class $\mathbb{S}_{\mathrm{NL}}$ [see Figure \ref{fig:EIV_NLTI_ExpValuePN_ClosedLoop}], where $\check{U}(k)=\mathbb{E}\{U(k)|r(t)\}$ and $\check{Y}(k)=\mathbb{E}\{Y(k)|r(t)\}$. The stochastic nonlinear distortion $Y_{\mathrm{S}}(k)$ is uncorrelated with -- but not independent of -- the reference $R(k)$.}
\label{fig:BLA_ExpVal_NLTI_PN_ClosedLoop}
\end{figure}

From Lemma \ref{lem:ExpValue_class_Sw} it follows that the new nonlinear plant in Figure \ref{fig:EIV_NLTI_ExpValuePN_ClosedLoop} can be replaced by its best linear approximation and an output residual (see Figure \ref{fig:BLA_ExpVal_NLTI_PN_ClosedLoop})
\begin{equation}
Y_{\mathrm{S}}(k) = \mathbb{E}\{Y(k)|r(t)\}-G_{\mathrm{BLA},N}(j\omega_k)\mathbb{E}\{U(k)|r(t)\}\label{eq:YS_ExpVal}
\end{equation}
that satisfy Properties \ref{enum:prop2_BLA} and \ref{enum:prop_YS} of the theorem [proof: the conditions of Theorem 3.22 on page 94 of \cite{Pintelon_book_2012} are fulfilled]. Property \ref{enum:prop1_BLA} guarantees that the BLAs in Figures \ref{fig:BLA_NLTI_PN_ClosedLoop} and \ref{fig:BLA_ExpVal_NLTI_PN_ClosedLoop} are the same, and it can easily be verified that \eqref{eq:YS_ExpVal} is identical to \eqref{eq:Def_YS}.

Following the same lines of the proof of Property \ref{enum:prop1_BLA}, the expected values in \eqref{eq:prop_YS_YP} are calculated using \eqref{eq:SplitExpValue}. We find
\begin{align}
\mathbb{E}\{Y_{\mathrm{S}}(k)\overline{Y_{\mathrm{P}}(l)}\}&=\mathbb{E}\{\mathbb{E}\{Y_{\mathrm{S}}(k)\overline{Y_{\mathrm{P}}(l)}\bigl|r(t)\}\}\nonumber\\
&=\mathbb{E}\{Y_{\mathrm{S}}(k)\mathbb{E}\{\overline{Y_{\mathrm{P}}(l)}\bigl|r(t)\}\}\nonumber\\
&=0\nonumber
\end{align}
where the second equality results from the fact that $Y_{\mathrm{S}}(k)$ is fixed for a given $r(t)$ [combine \eqref{eq:Def_Zs_PN}, \eqref{eq:ZBLA_R} and \eqref{eq:Def_YS}], and where the last equality uses $\mathbb{E}\{Y_{\mathrm{P}}(l)\}=0$ [combine \eqref{eq:Def_Zp_PN} and \eqref{eq:Def_YP}].

Since the nonlinear feedback system belongs to the class $\mathbb{S}_{\mathrm{NL},w}$ [see Definition \ref{def:NL_TI_systems_PN}], it is a two-input $x(t)=[r(t)\,w(t)]^T$, two-output $z(t)=[y(t)\,u(t)]^T$ version of the class $\mathbb{S}_{\mathrm{NL}}$ [see Definition \ref{def:NL_TI_systems}]. Therefore, the residual $Z_{\mathrm{res}}(k)=Z(k)-Z_{\mathrm{BLA}}(k)$ satisfies Property \ref{enum:prop_YS} of the theorem [proof: the conditions of Theorem 3.16 on page 86 of \cite{Pintelon_book_2012} are fulfilled]. Since $Z_{\mathrm{res}}(k)=\tilde{Z}_{\mathrm{S}}(k)+\tilde{Z}_{\mathrm{P}}(k)$ [see \eqref{eq:Zsplit}], it follows from \eqref{eq:Def_YS} and \eqref{eq:Def_YP} that
\begin{equation}
Y_{\mathrm{res}}(k)-G_{\mathrm{BLA},N}(j\omega_k)U_{\mathrm{res}}(k) = Y_{\mathrm{S}}(k)+Y_{\mathrm{P}}(k)
\label{eq:Yres_Ures_YP_YS}
\end{equation}
Given that $Y_{\mathrm{S}}(k)$ and $Y_{\mathrm{P}}(k)$ are mutually uncorrelated (Property \ref{enum:prop_YS_YP} of the theorem), that $Y_{\mathrm{res}}(k)$, $U_{\mathrm{res}}(k)$ and $Y_{\mathrm{S}}(k)$ all satisfy Property \ref{enum:prop_YS} of the theorem, and relationship \eqref{eq:Yres_Ures_YP_YS}, it can easily be shown  that $Y_{\mathrm{P}}(k)$   also satisfies Property \ref{enum:prop_YS}. For example, taking the expected value of the square of \eqref{eq:Yres_Ures_YP_YS}, using \eqref{eq:YS_CircCompl} and \eqref{eq:prop_YS_YP}, gives
\begin{equation}
O(N^{-1}) = O(N^{-1}) + \mathbb{E}\{Y_{\mathrm{P}}^2(k)\}
\end{equation}
which proves Property \ref{subenum:YS_CircCompl} of $Y_{\mathrm{P}}(k)$. The other properties are proven in exactly the same way.

\section{Step-Invariant Transform of the Volterra Kernels}
\label{sec:StepInvTransfVolterra}
First, we handle Volterra kernels of degree one and two. Next, the results are generalized to higher degree kernels.

The first degree Volterra kernel corresponds to the impulse response of an LTI system and its discretization is handled in standard text books [see, for example, \cite{MiddletonGoodwin_book_1990}]. Consider a continuous-time Volterra kernel of degree one with impulse response $g_1(t)$ excited by a piecewise constant input $u(t)$
\begin{subequations}
\label{subeq:ZOH_input}
\begin{equation}
u(t) = \sum_{n=-\infty}^{+\infty}u_n\mathrm{zoh}(t-nT_{\mathrm{s}})
\label{eq:ZOH_input}
\end{equation}
where
\begin{equation}
\mathrm{zoh}(t)=\begin{cases}
1 & 0\leqslant t<T_{\mathrm{s}}\\
0 & \mathrm{elsewhere}
\end{cases}
\label{eq:zoh_function}
\end{equation}
\end{subequations}
and with $T_{\mathrm{s}}$ the sampling period. It is shown that the input-output samples of the linear continuous-time system with impulse response $g_1(t)$ are exactly related by the following linear discrete-time transfer function
\begin{equation}
G_{1,\mathrm{zoh}}(z^{-1}) = (1-z^{-1})\mathcal{Z}\Bigl\{\mathcal{L}^{-1}\Bigl\{\frac{G_1(s)}{s}\Bigr\}\Bigr\}
\label{eq:StepInvariantTransf_FirstDegree}
\end{equation}
where $G_1(s)=\mathcal{L}\{g_1(t)\}$, with $\mathcal{L}\{\}$ the Laplace transform, $\mathcal{L}^{-1}\{\}$ the inverse Laplace transform, and $\mathcal{Z}\{\}$ the Z-transform of the sampled signal. Equation \eqref{eq:StepInvariantTransf_FirstDegree} is called the step-invariant transform of the continuous-time transfer function $G_1(s)$.

The response $y_2(t)$ of a second degree Volterra kernel to the input $u(t)$ \eqref{subeq:ZOH_input} is given by [use \eqref{eq:ConvKernelDegreeAlpha}]
\begin{equation}
y_2(t) = \int_0^{\infty}\int_0^{\infty}g_2(\tau_1,\tau_2)u(t-\tau_1)u(t-\tau_2)d\tau_1 d\tau_2
\label{eq:RespSecondDegreeVolterra}
\end{equation}
Sampling \eqref{eq:RespSecondDegreeVolterra} at $t=lT_{\mathrm{s}}$, taking into account \eqref{subeq:ZOH_input}, we find
\begin{align}
y_2(lT_{\mathrm{s}}) = &\sum_{n_1,n_2=-\infty}^{+\infty}u_{n_1}u_{n_2}\nonumber\\
&\int_{(l-n_1-1)T_{\mathrm{s}}}^{(l-n_1)T_{\mathrm{s}}}\int_{(l-n_2-1)T_{\mathrm{s}}}^{(l-n_2)T_{\mathrm{s}}}g_2(\tau_1,\tau_2)d\tau_1d\tau_2
\label{eq:SampledRespSecondDegreeVolterra}
\end{align}
Introducing the intermediate function
\begin{equation}
h_2(t_1,t_2) = \int_{0}^{t_1}\int_{0}^{t_2} g_2(\tau_1,\tau_2)d\tau_1 d\tau_2
\label{eq:DefIntermFunction}
\end{equation}
the sampled reponse \eqref{eq:SampledRespSecondDegreeVolterra} can be rewritten as
\begin{equation}
y_2(lT_{\mathrm{s}}) = \sum_{n_1,n_2=-\infty}^{+\infty}g_{2,\mathrm{zoh}}(l-n_1,l-n_2)u_{n_1}u_{n_2}
\label{eq:RespDiscreteTimeSecondDegreeVolterra}
\end{equation}
where
\begin{align}
&g_{2,\mathrm{zoh}}(n_1,n_2) = h_2(n_1T_{\mathrm{s}},n_2T_{\mathrm{s}})\nonumber\\
&+ h_2((n_1-1)T_{\mathrm{s}},(n_2-1)T_{\mathrm{s}})-h_2(n_1T_{\mathrm{s}},(n_2-1)T_{\mathrm{s}})\nonumber\\
&-h_2((n_1-1)T_{\mathrm{s}},n_2T_{\mathrm{s}})
\label{eq:DiscreteTimeSecondDegreeKernel}
\end{align}
It proves that the input-output samples of the continuous-time Volterra kernel of degree two are exactly related by a discrete-time Volterra kernel of degree two. Taking the two-dimensional $Z$-transform of \eqref{eq:DiscreteTimeSecondDegreeKernel} gives the following relationship between the two-dimensional discrete-time $G_{2,\mathrm{zoh}}(z_1^{-1},z_2^{-1})=\mathcal{Z}\{g_{2,\mathrm{zoh}}(n_1,n_2)\}$ and continuous-time $G_{2}(s_1,s_2)=\mathcal{L}\{g_{2}(\tau_1,\tau_2)\}$ transfer functions of the second degree Volterra kernels
\begin{align}
G_{2,\mathrm{zoh}}(z_1^{-1},z_2^{-1}) = &(1-z_1^{-1})(1-z_2^{-1})\nonumber\\
&\mathcal{Z}\Bigl\{\mathcal{L}^{-1}\Bigl\{\frac{G_2(s_1,s_2)}{s_1s_2}\Bigr\}\Bigr\}
\label{eq:StepInvTranfSecondDegreeKernel}
\end{align}
with $\mathcal{L}\{\}$ and $\mathcal{L}^{-1}\{\}$ the two-dimensional Laplace and inverse Laplace transforms, respectively. Equation \eqref{eq:StepInvTranfSecondDegreeKernel} is the step-invariant transform of the two-dimensional continuous-time transfer function $G_2(s_1,s_2)$.

Generalization of results \eqref{eq:StepInvariantTransf_FirstDegree} and \eqref{eq:StepInvTranfSecondDegreeKernel} to a Volterra kernel of degree $\alpha$ is straightforward
\begin{align}
G_{\alpha,\mathrm{zoh}}(z_1^{-1},\ldots,z_{\alpha}^{-1}) = &\prod_{i=1}^{\alpha}(1-z_i^{-1})\nonumber\\
&\mathcal{Z}\Bigl\{\mathcal{L}^{-1}\Bigl\{\frac{G_{\alpha}(s_1,\ldots,s_{\alpha})}{\prod_{i=1}^{\alpha}s_i}\Bigr\}\Bigr\}
\label{eq:StepInvTranfKernelDegreeAlpha}
\end{align}
which concludes the proof.

\section{Proof of Constraints \eqref{subeq:Constraints_alpha} and Equations \eqref{subeq:BLA_feedback_NFIR_PN}}
\label{sec:Proof_BLA_Feedback_NFIR_PN}
\subsection{Calculation of the Best Linear Approximation of \eqref{eq:NFIR_IO}}
\label{subsec:BLA_NFIR}
To calculate the BLA, we take the expected value of \eqref{subeq:Feedback_NFIR_PN}, given the reference signal $r(t)$. Using the notation
\begin{equation}
\check{x}(t)=\mathbb{E}\{x(t)|r(t)\}
\label{eq:CondExpValue}
\end{equation}
with $x=u,y$, and the independence of $u(t-2)$ and $w^2(t)$, we find
\begin{subequations}
\label{subeq:ExpValue_Feedback_NFIR}
\begin{align}
\check{y}(t) &= \check{u}(t-1) + \check{u}(t-2)\sigma_w^2\label{eq:ExpValue_IO}\\
\check{u}(t) &= r(t) -\alpha\check{y}(t)\label{eq:ExpValue_Feedback}
\end{align}
\end{subequations}
which corresponds to a linear time-invariant system. Elimination of $\check{u}$ in \eqref{subeq:ExpValue_Feedback_NFIR}, gives the following linear time-invariant relationship between $r(t)$ and $\check{y}(t)$
\begin{equation}
\check{y}(t) + \alpha\check{y}(t-1) + \alpha\sigma_w^2\check{y}(t-2) = r(t-1) + \sigma_w^2r(t-2)
\label{eq:Rel_ExpValue_y_r}
\end{equation}
Hence, the BLA from reference $r(t)$ to output $y(t)$ equals
\begin{equation}
G_{\mathrm{BLA},ry}(z^{-1}) = \frac{z^{-1} + \sigma_w^2 z^{-2}}{1 + \alpha z^{-1} + \alpha\sigma_w^2 z^{-2}}
\label{eq:BLA_ry_feedback_NFIR_PN}
\end{equation}
[proof: take the $Z$-transform of \eqref{eq:Rel_ExpValue_y_r}]. Eliminating $\check{y}(t)$ in \eqref{subeq:ExpValue_Feedback_NFIR}, we find in a similar way the BLA from reference $r(t)$ to input $u(t)$
\begin{equation}
G_{\mathrm{BLA},ru}(z^{-1}) = \frac{1}{1 + \alpha z^{-1} + \alpha\sigma_w^2 z^{-2}}
\label{eq:BLA_ru_feedback_NFIR_PN}
\end{equation}
Dividing \eqref{eq:BLA_ry_feedback_NFIR_PN} by \eqref{eq:BLA_ru_feedback_NFIR_PN} gives the BLA from input $u(t)$ to output $y(t)$
\begin{equation}
G_{\mathrm{BLA}}(z^{-1}) = \frac{G_{\mathrm{BLA},ry}(z^{-1})}{G_{\mathrm{BLA},ru}(z^{-1})} = z^{-1} + \sigma_w^2 z^{-2}
\label{eq:Gbla_Gblayr_div_Gblaru}
\end{equation}
which proves \eqref{eq:BLA_feedback_NFIR_PN}.
\subsection{Derivation of the Stability Constraints \eqref{subeq:Constraints_alpha}}
\label{subsec:Constr_alpha}
Imposing that the poles of the BLAs \eqref{eq:BLA_ry_feedback_NFIR_PN} and \eqref{eq:BLA_ru_feedback_NFIR_PN}
\begin{equation}
z = 0.5(-\alpha \pm \sqrt{\alpha^2 - 4\alpha\sigma_w^2})
\label{eq:PolesBLAs}
\end{equation}
are complex conjugate ($\alpha^2<4\alpha\sigma_w^2$), results in the constraint
\begin{equation}
0<\alpha<4\sigma_w^2
\label{eq:ComplexPoles}
\end{equation}
Stability of the poles \eqref{eq:PolesBLAs} satisfying \eqref{eq:ComplexPoles}, requires that
\begin{equation}
|z|^2<1~~\Rightarrow~~\alpha < \sigma_w^{-2}
\label{eq:StabPoles}
\end{equation}
Combining  \eqref{eq:ComplexPoles} and \eqref{eq:StabPoles} proves \eqref{eq:NFIR_IO}.

If $\sigma_w=0$, then the pole of the BLAs \eqref{eq:BLA_ry_feedback_NFIR_PN} and \eqref{eq:BLA_ru_feedback_NFIR_PN} equals $z=-\alpha$, which shows \eqref{eq:Constraint_alpha_zero_PN}.

\subsection{Nonlinear Distortion \eqref{eq:ys_feedback_NFIR_PN} and Process Noise \eqref{eq:yp_feedback_NFIR_PN}}
\label{subsec:YS_YP}
\begin{figure}[t!] %[htbp]: here, top, bottom, page (follow by ! to enforce, e.g. b!)
\centering
\includegraphics[width=0.49\columnwidth]{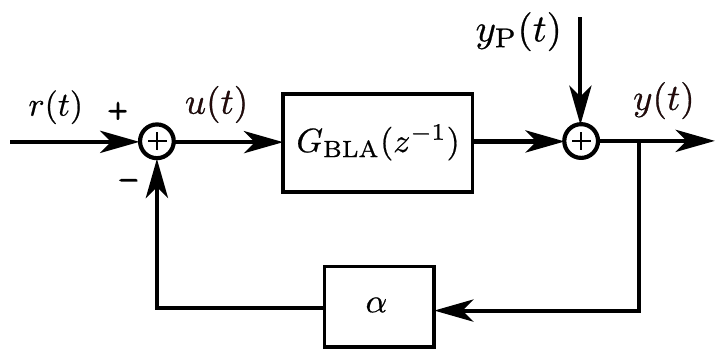}
\caption{Best linear approximation \eqref{eq:Gbla_Gblayr_div_Gblaru} and process noise \eqref{eq:Deriv_yp} of the closed loop NFIR system \eqref{subeq:Feedback_NFIR_PN}.}
\label{fig:BLA_CL_NFIR_PN}
\end{figure}

Since equations \eqref{subeq:ExpValue_Feedback_NFIR} are linear, the stochastic nonlinear distortion is zero. Using \eqref{eq:Gbla_Gblayr_div_Gblaru} and taking into account that $y_{\mathrm{s}}(t)=0$, we find,
\begin{align}
y_{\mathrm{P}}(t) & =y(t)-G_{\mathrm{BLA}}(q)u(t)\nonumber \\
 & =u(t-2)[w^{2}(t)-\sigma_{w}^{2}]\label{eq:Deriv_yp}
\end{align}
where $q$ is the backward shift operator [$qx(t)=x(t-1)$]. This leads to the block diagram shown in Figure \ref{fig:BLA_CL_NFIR_PN}.
\subsection{Variance of the BLA estimate \eqref{eq:varBLA_feedback_NFIR_PN}}
\label{subsec:var_BLS}
First, the variance of the BLA estimate is calculated assuming that the closed loop NFIR system \eqref{subeq:Feedback_NFIR_PN} operates under periodic steady state. Next, it is shown that the variance \eqref{eq:varBLA_feedback_NFIR_PN} is independent of $\sigma_r^2$. Finally, the connection with the local polynomial estimate from the transient response to a random phase multisine excitation $r(t)$ is established.

Under the periodic state state assumption, the input-output DFT spectra $U(k)$ and $Y(k)$ are related to the periodic reference $R(k)$ and the process noise $Y_{\mathrm{P}}(k)$ as
\begin{subequations}
\label{subeq:IO_Ref_ProcNoise}
\begingroup
\allowdisplaybreaks
\begin{align}
U(k) & =\frac{R(k)}{1+\alpha G_{\mathrm{BLA}}(j\omega_{k})}-\frac{\alpha Y_{\mathrm{P}}(k)}{1+\alpha G_{\mathrm{BLA}}(j\omega_{k})}\label{eq:Input_RefProcNoise}\\
Y(k) & =\frac{G_{\mathrm{BLA}}(j\omega_{k})R(k)}{1+\alpha G_{\mathrm{BLA}}(j\omega_{k})}+\frac{Y_{\mathrm{P}}(k)}{1+\alpha G_{\mathrm{BLA}}(j\omega_{k})}\label{eq:Output_RefProcNoise}
\end{align}
\endgroup
\end{subequations}
From \eqref{subeq:IO_Ref_ProcNoise} follow the input-output (co-)variances, given the reference $R(k)$
\begin{subequations}
\label{subeq:IO_covar_ProcNoise}
\begingroup
\allowdisplaybreaks
\begin{align}
\sigma_{U}^{2}(k) &= \frac{\alpha^{2}\sigma_{\mathrm{P}}^{2}(k)}{|1+\alpha G_{\mathrm{BLA}}(j\omega_{k})|^{2}}\label{eq:I_variance_PN}\\
\sigma_{Y}^{2}(k) &=\frac{\sigma_{\mathrm{P}}^{2}(k)}{|1+\alpha G_{\mathrm{BLA}}(j\omega_{k})|^{2}}\label{eq:O_variance_PN}\\
\sigma_{YU}^{2}(k) &= \frac{-\alpha\sigma_{\mathrm{P}}^{2}(k)}{|1+\alpha G_{\mathrm{BLA}}(j\omega_{k})|^{2}}\label{eq:IO_covariance_PN}
\end{align}
\endgroup
\end{subequations}
with $\sigma_{\mathrm{P}}^{2}(k)=\mathrm{var}(Y_{\mathrm{P}}(k))$.

For one realization of the random phase multisine excitation $r(t)$, the spectral analysis definition \eqref{eq:G_BLA_PN_N} of the BLA simplifies to
\begin{equation}
\hat{G}_{\mathrm{BLA}}(j\omega_k) = \frac{Y(k)}{U(k)}
\label{eq:Gbla_Estim_OneRealization}
\end{equation}
The variance of the BLA estimate \eqref{eq:Gbla_Estim_OneRealization} can be approximated as
\begin{align}
\mathrm{var}(\hat{G}_{\mathrm{BLA}}(j\omega_{k}))\approx|&G_{\mathrm{BLA}}(j\omega_{k})|^{2}\Bigl(\frac{\sigma_{Y}^{2}(k)}{|Y_{0}(k)|^{2}}+\frac{\sigma_{U}^{2}(k)}{|U_{0}(k)|^{2}}\nonumber\\
&-2\mathrm{Re}\bigl(\frac{\sigma_{YU}^{2}(k)}{Y_{0}(k)\overline{U_{0}(k)}}\bigr)\Bigr)\label{eq:Var_BLA}
\end{align}
where $U_0(k)$ and $Y_0(k)$ are the parts of $U(k)$ and $Y(k)$ depending on $R(k)$ [see \cite{Pintelon_book_2012}, Section 2.4, pages 44--47]. Combining \eqref{subeq:IO_covar_ProcNoise} and \eqref{eq:Var_BLA} gives
\begin{equation}
\mathrm{var}(\hat{G}_{\mathrm{BLA}}(j\omega_{k}))\approx |1+\alpha G_{\mathrm{BLA}}(j\omega_{k})|^{2}\frac{\sigma_{\mathrm{P}}^{2}(k)}{|R(k)|^2}\label{eq:Var_BLA_contnd}
\end{equation}
Approximating $y_{\mathrm{P}}(t)$ by a white noise process, taking into account that $|R(k)|$ is independent of $k$, allows one to simplify the ratio in \eqref{eq:Var_BLA_contnd}
\begin{equation}
\frac{\sigma_{\mathrm{P}}^{2}(k)}{|R(k)|^2} \approx \frac{\mathrm{var}(y_{\mathrm{P}}(t))}{\mathrm{var}(r(t))}
\label{eq:Ratio_in_varGbla}
\end{equation}
Since $u(t-2)$ is independent of $w(t)$ and since $w(t)$ is normally distributed, the variance of $y_{\mathrm{P}}(t)$ \eqref{eq:Deriv_yp} equals
\begin{equation}
\mathrm{var}(y_{\mathrm{P}}(t)) = \sigma_u^2 2\sigma_w^4
\label{eq:Var_yp}
\end{equation}
Combining \eqref{eq:Var_BLA_contnd}, \eqref{eq:Ratio_in_varGbla} and \eqref{eq:Var_yp} proves \eqref{eq:varBLA_feedback_NFIR_PN}.

Eliminating $y(t)$ in \eqref{subeq:Feedback_NFIR_PN}, gives
\begin{equation}
u(t) + \alpha u(t-1) + \alpha w^2(t) u(t-2) = r(t)
\label{eq:Rel_u_r_CL_NFIR_PN}
\end{equation}
Multiplying both sides of \eqref{eq:Rel_u_r_CL_NFIR_PN} by $\beta\ne 0$ shows that $\beta u(t)$ is response to $\beta r(t)$. Hence, the ratio $\sigma_u^2/\sigma_r^2$ in \eqref{eq:varBLA_feedback_NFIR_PN} is independent of $\sigma_r^2$.

Compared with the BLA estimate \eqref{eq:Gbla_Estim_OneRealization}, the local polynomial estimate of the BLA reduces the estimation variance with a factor equal to the difference between the local number of frequencies used for the polynomial approximation and the number of local parameters. This difference is called the degrees of freedom $\mathrm{dof}$.

\end{document}